\documentclass[conference]{IEEEtran}
\IEEEoverridecommandlockouts
\usepackage{cite}
\usepackage{amsmath,amssymb,amsfonts}
\usepackage{hyperref}
\hypersetup{
  colorlinks   = true,		
  urlcolor     = blue,    	
  linkcolor    = blue,    	
  citecolor    = blue      	
}
\usepackage[english]{babel}
\usepackage{enumitem}
\usepackage{algorithmic}
\usepackage{float}
\usepackage{graphicx}
\usepackage{textcomp}
\usepackage{xcolor}
\usepackage{comment}
\usepackage[]{placeins}
\def\BibTeX{{\rm B\kern-.05em{\sc i\kern-.025em b}\kern-.08em T\kern-.1667em\lower.7ex\hbox{E}\kern-.125emX}}
\newcommand{\rot}[1]{{\ \textcolor{red}{#1}}}

\begin{document}

\title{Quantum fluctuations and lineshape anomaly in a high-$\beta$ silver-coated InP-based metallic nanolaser}

\author{\IEEEauthorblockN{
 A. Koulas-Simos\textsuperscript{1\rot{*}},  J. Buchgeister\textsuperscript{3\rot{*}}, M. Drechsler\textsuperscript{3\rot{*}}, T. Zhang\textsuperscript{2}, K. Laiho\textsuperscript{1, \#}, G. Sinatkas\textsuperscript{1}, J. Xu\textsuperscript{2},
 F. Lohof\textsuperscript{3}, \\ Q. Kan\textsuperscript{2}, R. K. Zhang\textsuperscript{2}, F. Jahnke\textsuperscript{3}, C. Gies\textsuperscript{3}, W. W. Chow\textsuperscript{4},
 C. Z. Ning\textsuperscript{2,5+} and S. Reitzenstein\textsuperscript{1+}
}
\IEEEauthorblockA{\textit{\textsuperscript{1}Institut f{\"u}r Festk\"orperphysik, Technische Universit{\"a}t Berlin,
Hardenbergstr. 36, 10623 Berlin, Germany.} \\
\textit{\textsuperscript{2}Department of Electronic Engineering, Tsinghua University, 1303 Beijing 100084, China.} \\
\textit{\textsuperscript{3}Institut f{\"u}r Theoretische Physik, Universit{\"a}t Bremen,
Otto-Hahn-Allee 1, 28359 Bremen, Germany.} \\
\textit{\textsuperscript{4}Sandia National Laboratories, Albuquerque, New Mexico, USA.} \\
\textit{\textsuperscript{5}School of Electrical, Computer and Energy Engineering, Arizona State University, 650 E Tyler Mall, AZ 85281, USA.} \\
\textsuperscript{\rot{*}}These authors contributed equally to this work.\\
 \textsuperscript{\#} present address: Institute of Quantum Technologies, German Aerospace Center (DLR), Wilhelm-Runge-Str. 10,  89081 Ulm, Germany\\
 \textsuperscript{+} e-mail for correspondence: cning@mail.tsinghua.edu.cn, stephan.reitzenstein@physik.tu-berlin.de }}


\maketitle

\begin{abstract}
Metallic nanocavity lasers provide important technological advancement towards even smaller integrable light sources. They give access to widely unexplored lasing physics in which the distinction between different operational regimes, like those of thermal or a coherent light emission, becomes increasingly challenging upon approaching a device with a near-perfect spontaneous-emission coupling factor $\beta$. In fact, quantum-optical studies have to be employed to reveal a transition to coherent emission in the intensity fluctuation behavior of nanolasers when the input-output characteristic appears thresholdless for $\beta = 1$ nanolasers. Here, we identify a new indicator for lasing operation in high-$\beta$ lasers by showing that stimulated emission can give rise to a lineshape anomaly manifesting as a transition from a Lorentzian to a Gaussian component in the emission linewidth that dominates the spectrum above the lasing threshold.
\end{abstract}


\section{Introduction}

The demand for energy-efficient, miniaturized, and integrable light-emitting devices is a strong driving force in optoelectronics
\cite{Miller2009,Ning2019} and quantum nanophotonics~\cite{Rodt2021}. Advanced low-threshold semiconductor lasers have resulted in significant technological achievements to reach more and more compact devices with enhanced lasing performance \cite{Wang2018,Jeong2020,Liang2020,Azzam2020,Deng2021}.
At the same time, the exploration of the quantum limit of lasing has brought up fundamental questions about the lasing threshold and its
identification in micro- and nanolasers \cite{Bjork1994,Ning2013,Strauf2006,Samuel2009,Chow2014,Kreinberg2017,Jagsch2018,Reeves2018,Wu2019}. In such lasers, the efficient coupling of the spontaneous emission of the gain material into the lasing cavity mode, expressed by the
spontaneous emission coupling factor $\beta$ close to the ideal value of 1, leads to a kink-free behavior in the input-output characteristics. While fluctuation of the emission in terms of the second-order photon autocorrelation function
are often taken into consideration for identifying the onset of lasing in high-$\beta$ nanolasers \cite{Chow2018}, here we introduce a new
indicator that can be simply extracted from an analysis of the emission lineshape alone.

From cavity quantum electrodynamics (cQED), the emission spectrum of a laser is generally understood to have a Lorentzian shape that narrows with increasing intracavity photon number, according to the Shawlow-Townes formula \cite{Schawlow1958}, or the modified one for lasers above threshold \cite{Scully1966,Haken1966,Lax1966,Haug1967}. For a strongly inhomogeneous gain medium, Gaussian inhomogeneous broadening is well known to give rise to a Gaussian emission lineshape in the low-excitation regime \cite{Glauser2014}. With increasing excitation strength the cavity acts like a spectral filter and promotes stimulated emission, singling out resonant transitions from the gain medium, leading to a Lorentzian lineshape. Here, we observe the opposite behavior, which is a transition from a Lorentzian to a Gaussian lineshape at the
lasing threshold. In the past, Gaussian spectral components in the lasing regime have been shown to result from $1/f$ noise like carrier density fluctuations \cite{Stephan2005,Septon2019}. In this paper, we provide a different description of the quasi (Fox-Li) eigenmodes of a cavity with outcoupling to show that the lineshape transition occurs at the threshold due to intrinsic factors arising from nonlinearities in the active medium.

The combined experimental and theoretical study presented here is centered around an InP-based silver-coated nanolaser emitting at telecom wavelength
in a low temperature environment of 10$\,$K. In contrast to dielectric cavity structures, plasmonic and metal-clad nanolasers are capable of deep
sub-wavelength physical volumes and cavity mode volumes below the fundamental size limit of the cubic half-wavelength \cite{Hill2009,Oulton2009,Noginov2009, Khajavikhan2012, Kwon2012}. Since their first demonstration, metal-cavity nanolasers have gained significant momentum over their dielectric counterparts by breaking the aforementioned fundamental size limit and leading to the first observation of  lasing without a kink in the input-output curve \cite{Khajavikhan2012}.

\section{Experimental Details}

\begin{figure*} [!h]
	\centering
	\includegraphics[width=0.8\linewidth]{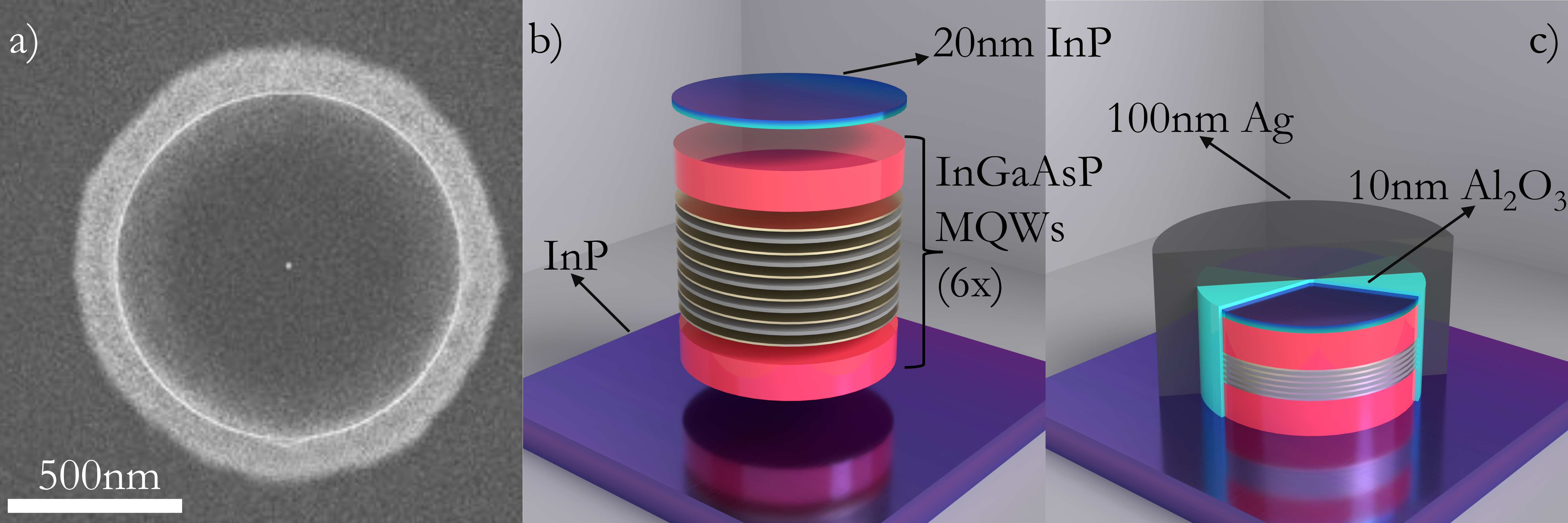}
	\caption{a) SEM image of an exemplary cylindrical MQW nanolaser device with a designed fabrication diameter of 700$\,$nm, a scale of 500$\,$nm
	width is provided in white. b) and c) Schematic representation of the the MQW composition highlighting the fabrication process: a 100$\,$nm layer
	of In$_{0.78}$Ga$_{0.22}$As$_{0.49}$P$_{0.51}$ is followed by 6 MQWs consisting of 6 layers of 6$\,$nm In$_{0.84}$Ga$_{0.16}$As$_{0.66}$P$_{0.34}$
	gain material and 10$\,$nm In$_{0.73}$Ga$_{0.27}$As$_{0.53}$P$_{0.47}$ barrier material, topped with another 100$\,$nm of the
	In$_{0.78}$Ga$_{0.22}$As$_{0.49}$P$_{0.51}$ as well as 20$\,$nm InP as a capping layer. The entire device is then encapsulated in a 10$\,$nm
	Al$_{2}$O$_{3}$ layer for optical loss insulation and a 100$\,$nm layer of Ag to realize the cavity. In a final step, the device is glued on a
	Si wafer from the Ag side, flipped 180$^\circ$ and has the InP base layer removed to achieve operability.}
	\label{fig:Sample}
\end{figure*}

Fig. \ref{fig:Sample} shows a scanning electron microscope (SEM) image of an exemplary nanolaser device with a diameter of 700$\,$nm as well as a schematic representation of the multiple quantum well (MQW) nanolaser design. The gain material consists of a total of 6 InGaAsP quantum wells grown on an InP substrate with the metallic cavity being realized via a 100$\,$nm thick silver capping, while the 10$\,$nm dielectric layer of
$\text{Al}_{2}\text{O}_3$ shields the structure from optical losses in metals at visible and NIR wavelengths through dissipation and
surface carrier recombination.

\begin{figure} [!t]
	\centering
	\includegraphics[width=1.0\linewidth]{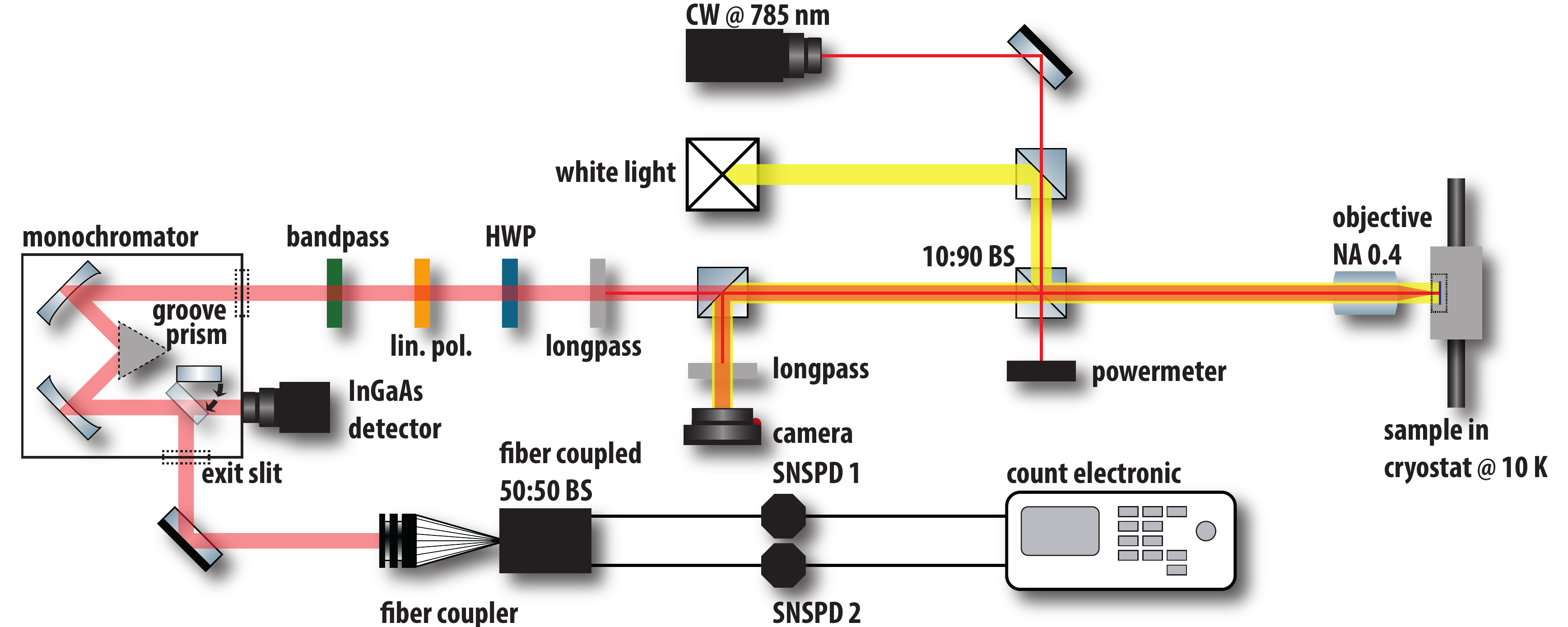}
	\caption{Schematic view of the experimental setup consisting of a high-resolution $\mu$PL setup in conjunction with a fiber-based HBT
	configuration for optical and quantum optical measurements.}
	\label{fig:Setup}
\end{figure}

The experimental configuration used for the optical and quantum optical study of the metallic nanolaser (MNL) is shown in Fig. \ref{fig:Setup}. The investigation relies on high-resolution micro-photoluminescence ($\mu$PL) spectroscopy with a spectral resolution of 0.05$\,$nm in conjunction with photon-autocorrelation measurements using a fiber-based Hanbury Brown and Twiss (HBT) configuration \cite{Brown1956} with a temporal resolution of 80$\,$ps under continuous-wave (CW) operation at 785$\,$nm. All measurements presented in this study have been performed on a selected MQW nanolaser with a diameter of 700$\,$nm. More details can be found in the section "Methods" or in Ref. \cite{Kreinberg2020}. 

\begin{figure} [!t]
	\centering
	\includegraphics[width=0.8\linewidth]{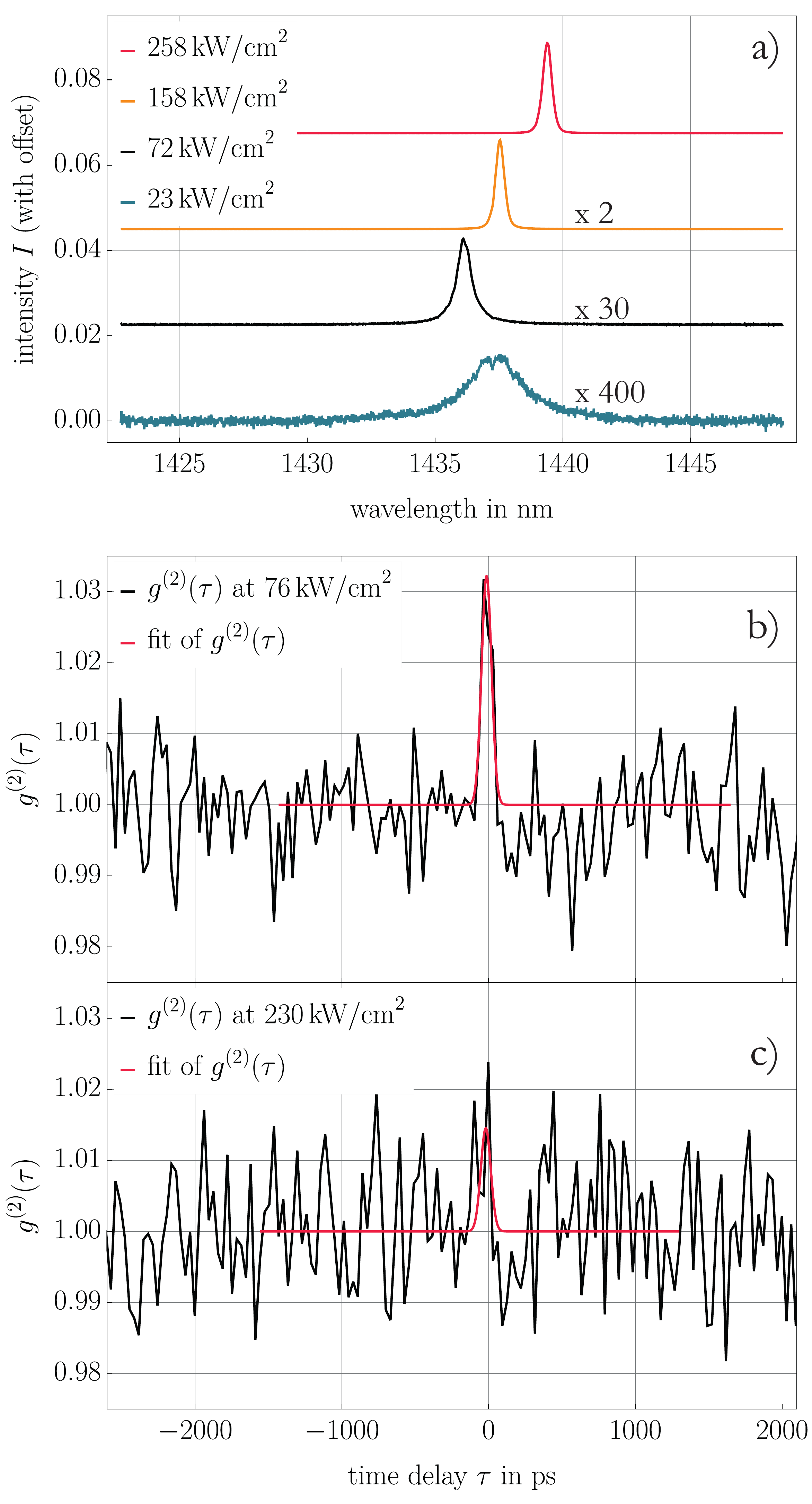}
	\caption{a) Recorded $\mu$PL spectra of the MQW nanolaser device at 10$\,$K for various excitation power densities between 23 and 258$\,\text{kW}\text{cm}^{-2}$. From lowest to highest excitation power-density, the spectra are multiplied with a factor of 400, 30, 2 and 1 respectively for a unified depiction. Panels b) and c) show measured autocorrelation traces for $g^{(2)}(\tau)$ for a pump rate near and well above the laser threshold, respectively.}
	\label{fig:Spectra}
\end{figure}

\begin{figure*} [!t]
	\centering
	\makebox[\textwidth][c]{
	\includegraphics[width=1.05\linewidth]{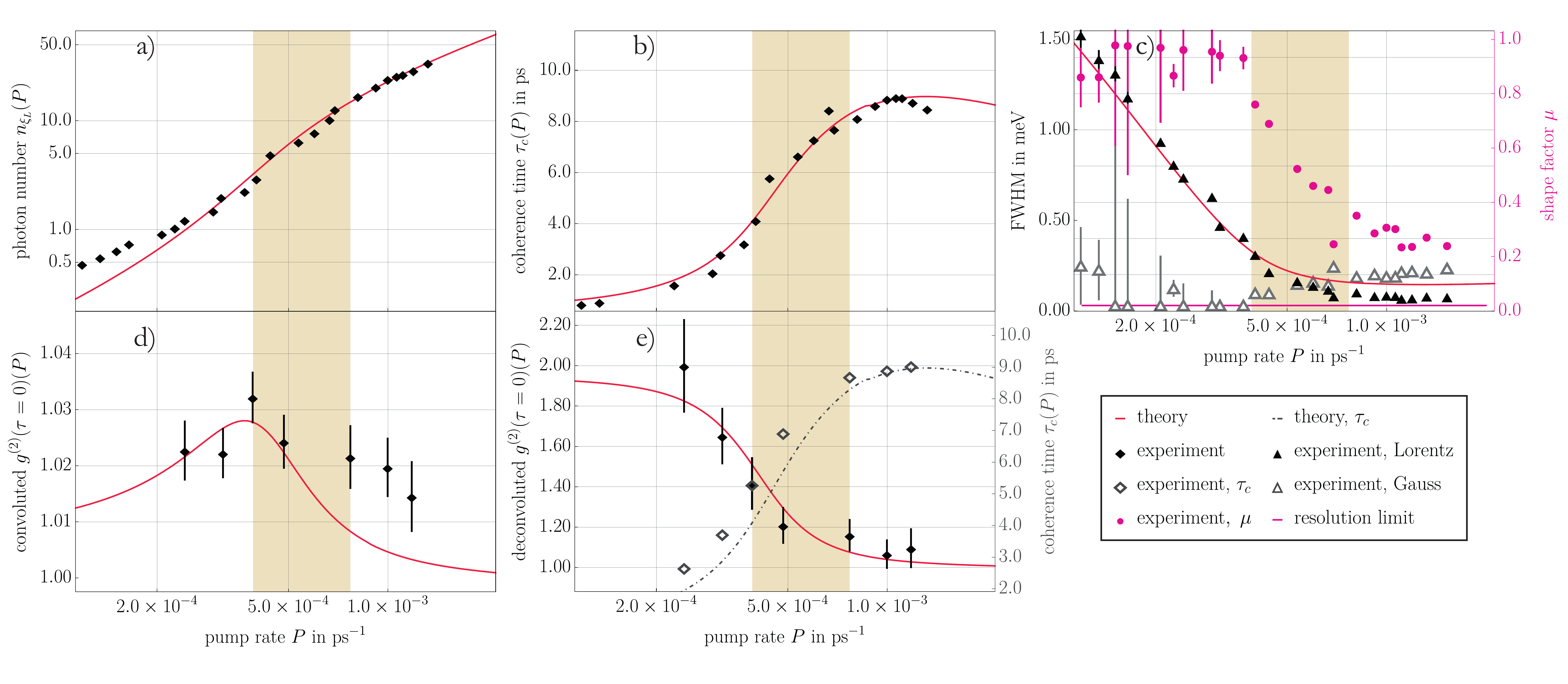}}
	\caption{Excitation-power dependent characterization of the MNL showing a comparison of experimental
	and theoretical data obtained from quantum-optical measurements and modelling, respectively. Namely: a) input-output characteristics, b) coherence time, and c) Gaussian and Lorentzian linewidth contributions together with the calculated shape factor obtained from a spectral Voigt analysis. Panels d) and e) show the second-order photon-autocorrelation function $g^{(2)}(0)$: in d) the raw experimental data is compared with the theoretical result that has been convoluted with a Gaussian setup response function, in e) the deconvoluted data clearly exhibits the transition to coherent light in agreement with the theoretical analysis.}
	\label{fig:Comparison}
\end{figure*}

\section{Quantum-optical characterization}

Excitation power dependent $\mu$PL measurements were conducted at 10$\,$K to explore emission properties in terms of a nonlinear input-output
power dependence of the emission intensity and linewidth narrowing above threshold. Fig. \ref{fig:Spectra}a) shows a set of the recorded emission spectra at excitation power densities ranging from 23 to 258$\,{\text{kW}}/{\text{cm}^2}$. A broad emission feature with a full width at half maximum (FWHM) of about 2.5$\,$nm is observed for low input powers, suggesting thermal emission in the spontaneous-emission regime. With increasing excitation power, a sharp emission line arises at a cavity mode energy, indicating a transition to coherent emission.

A microscopic semiconductor laser model is used to capture the excitation and emission dynamics of the MQW gain material embedded in the
nanocavity. Details of the theoretical modelling process are given in the accompanying SI. Our description uses the quantized light field and, therefore, naturally contains quantum fluctuations giving rise to spontaneous emission. Two-time calculations for the real time $t$ and the delay time $\tau$ are used to obtain $g^{(1)}(t,t+\tau)$, from which the coherence time $\tau_{\text{coh}}$ is calculated. Furthermore, the model gives access to the
second-order photon-autocorrelation function $g^{(2)}(0)$, allowing us to unambiguously distinguish lasing operation from thermal emission for our device.

The theory is evaluated for a single set of parameters with which we can reproduce the experimental results, which are both shown in Fig.~\ref{fig:Comparison}. The input-output characteristics are presented in Fig.~\ref{fig:Comparison}a), revealing a smooth s-shaped transition with shallow
threshold typical for a high-$\beta$ laser. The deviation of experimental and theoretical data at small pump rates is likely due to 0D-defects in the
active region of the device that contribute photons at the lasing wavelength \cite{Jagsch2018}; since this effect is subject to saturation, it becomes negligible once stimulated emission commences. Fig.~\ref{fig:Comparison}b) depicts the coherence time $\tau_{\text{coh}}$ with good agreement between theory and experiment. The prominent increase from 0.5$\,$ps to 9$\,$ps is indicative for the transition from spontaneous emission below to stimulated emission above threshold. The experimental coherence time data are extracted from the recorded spectra using a Voigt profile. In contrast to the previously used Pseudo-Voigt lineshape \cite{Kreinberg2020}, this allows us to establish a clear distinction between Lorentzian ($\gamma_{\text{L}}$) and Gaussian ($\gamma_{\text{G}}$) FWHM contributions to the overall lineshape shown in Fig.~\ref{fig:Comparison}c), the importance of which for identifying the onset of coherent emission will be discussed in greater detail in Section IV. From the Voigt fit we obtain the coherence time via

\begin{equation}
\tau_{\text{coh}}=\sqrt{\frac{2\ln2}{\pi\gamma^2_{\text{G}}}}\: 2^{\left(\frac{\gamma_{\text{L}}}{\gamma_{\text{G}}}\right)^2} \text{erfc}
\left(\sqrt{2\ln2}\frac{\gamma_{\text{L}}}{\gamma_{\text{G}}}\right).
\end{equation}

In the last decade, an increasing number of publications have established the importance of quantum optical studies on the emission
statistics to unambiguously prove lasing operation of high-$\beta$ emitters \cite{Strauf2006, Ulrich2007,Samuel2009,Wiersig2009,Hostein2010,Assmann2010, Hopfmann2013,Takiguchi2016,Ota2017,Kreinberg2017,Vyshnevyy2018,Jagsch2018,Reeves2018,Chow2018,Gies2019,Kreinberg2020}.
In fact, devices working in the regime of amplified spontaneous emission can exhibit linear input-output characteristics without a pronounced kink
and significant linewidth narrowing/coherence time increase, which could incorrectly be interpreted as a signature of lasing in a high-$\beta$ device \cite{Kreinberg2017}. These devices, however, do \emph{not} enter the coherent emission regime of a laser, which is only evidenced by accessing the statistical properties of the emission in quantum optical measurements.

To validate lasing in our device, we first performed an analysis in terms of the autocorrelation function $g^{(2)}(\tau)$, which is shown in Fig.~\ref{fig:Spectra}b) and c) for pump rates below and above the laser threshold. The transition from thermal to coherent emission is expected to manifest itself in the zero-time-delay value as a reduction of pronounced bunching with a normalized peak height of $g^{(2)}(0)=2$ to the Poisson level of $g^{(2)}(0)=1$. In order to directly compare experimental and numerical results for $g^{(2)}(0)$, an additional step is required. Due to the short coherence times of few ps, the raw data is strongly convolved with the temporal resolution (80$\,$ps) of the HBT detection setup, resulting in the data points shown in  Fig.~\ref{fig:Comparison}d), which do not show the expected transition from 2 to 1. The bunching effect is significantly suppressed -- an experimental issue which has already been reported in previous works \cite{Strauf2006,Ulrich2007,Reitzenstein2008a,Redlich2016,Kreinberg2020}.
In the low-excitation regime, the coherence time is too short for resolving the thermal component of $g^{(2)}(\tau)$. As the excitation power increases upon reaching the lasing threshold, the coherence time of the device increases, and partially coherent light (containing contributions of thermal and coherent light) appears, leading to a maximum of observed photon bunching with values of $g^{(2)}(0)=1.033$. A further increase of the excitation power results in $g^{(2)}(0)=1$ as expected for coherent light. We apply an approach based on the Siegert relation \cite{Kreinberg2020} to extract the deconvoluted $g^{(2)}(0)$, which is shown in Fig.~\ref{fig:Comparison}e), which reaches values of 1.99 at low powers
and transitions to values lower than 1.10 with increasing power in good agreement with the numerical results that natively give the deconvoluted $g^{(2)}(0)$. We note that a slight shift of the observed characteristics to lower pump rates is visible for the experimental photon-autocorrelation measurements, signifying higher effective power density (due to better optical adjustment) in this case. As a result, the power-dependent curves in Fig.~~\ref{fig:Comparison}a)--c) extracted from the optical studies are artificially shifted to higher powers, which is evident in the comparison of the average coherence time extracted from the spectra prior and after every $g^{(2)}$-measurement [shown in Fig.~\ref{fig:Comparison}e)], and the one extracted from the optical measurements in Fig.~~\ref{fig:Comparison}b).

We have also reversed the above-described procedure and performed a convolution of the calculated $g^{(2)}(0)$, which is added to  Fig.~\ref{fig:Comparison}d) in good agreement with the measurement. In conclusion, altogether the results of the quantum-optical investigations give clear evidence that our device indeed operates in the lasing regime.

\section{Anomalous threshold behavior of the laser lineshape}

The observation and explanation of the transition to a dominating Gaussian lineshape component at the laser threshold is a key finding of this work.
In previous publications, the lineshape of the spectrum was investigated using a Pseudo-Voigt profile, which allowed a simple analysis of the Lorentzian
and Gaussian contribution by means of a single shape factor \cite{Kreinberg2020}. In order to distinguish between the individual contributions, however, a full Voigt profile is better suited to analyze the spectra, since it offers the possibility to obtain separately the Lorentzian and the Gaussian linewidth contributions. Moreover, it provides access to the shape factor $\mu$ of the Voigt line profile approximately via $\mu=\frac{\gamma_{\text{L}}}{\gamma_{\text{L}}+\gamma_{\text{G}}}$, which is the weighted ratio of the Lorentzian ($\mu=1$) and Gaussian ($\mu=0$) linewidths. In Fig. \ref{fig:Comparison}c) the FWHM of both components are shown together with the shape factor $\mu$. At low excitation powers, the Lorentzian lineshape completely dominates the emission spectra and decreases about 8-fold with increasing excitation power. The clear Lorentzian lineshape below threshold can be attributed to the spontaneous emission and amplified spontaneous emission of the MQW gain material coupled into the resonator and is well described by considering cQED effects (red line). At the pump rate of $P \approx 5 \cdot 10^{-4}\,$ps$^{-1}$, the Gaussian component becomes comparable with the Lorentzian one. For it, we find a linewidth that approximately remains constant even at higher excitation powers, while the Lorentzian contribution further decreases but stays above the resolution limit (0.03$\,$meV) of the setup, leading to an overall Gaussian lineshape above threshold. This lineshape anomaly is clearly observed in the power-dependent behavior of the shape factor that drops from values near 1 close to the threshold pumping rate $P \approx 3 \cdot 10^{-4}\,$ps$^{-1}$ to 0.3, reflecting in a straightforward manner the dominance of the Gaussian component. While a similar behavior has previously been reported \cite{Kreinberg2020}, no explanation could be given for this deviation from the expected Lorentzian lineshape so far. At the beginning of this transition, the linewidth seems to have reached the lowest limit of about 0.2 meV and the central wavelength starts to redshift [see Fig.~\ref{fig:Spectra}a)], which could indicate heating-induced inhomogeneous broadening \cite{Kreinberg2020}. Another possible explanation could be the slightly different light-matter interaction-strengths of the individual QWs and the lasing mode due to the position-dependent overlap of the electronic wave functions and the lasing mode. In the SI we provide results from additional numerical calculations, showing that such effects actually cause no changes in the lineshape behavior above threshold, but lead to a Gaussian component in the low-excitation regime. Moreover, previous works \cite{Stephan2005} have attributed the Gaussian part to technical noise, such as charge density fluctuations, which could be associated with small mechanical instabilities in the cryostat leading to fluctuations of the excitation power density of the nanolasers. This would have led to the same constant lower resolution limit for the recorded linewidth of every investigated device in this setup. However, such extrinsic contributions have neither been observed in past conducted studies \cite{Kreinberg2020} and ours using the same setup, given in addition the fact that the linewidths are not resolution limited ($>$0.03$\,$meV) and, therefore, they can be ruled out. In contrast, our theoretical calculations show intrinsic effects leading to the experimentally obtained results.

\begin{figure} [!t]
	\centering
	\includegraphics[width=0.8\linewidth]{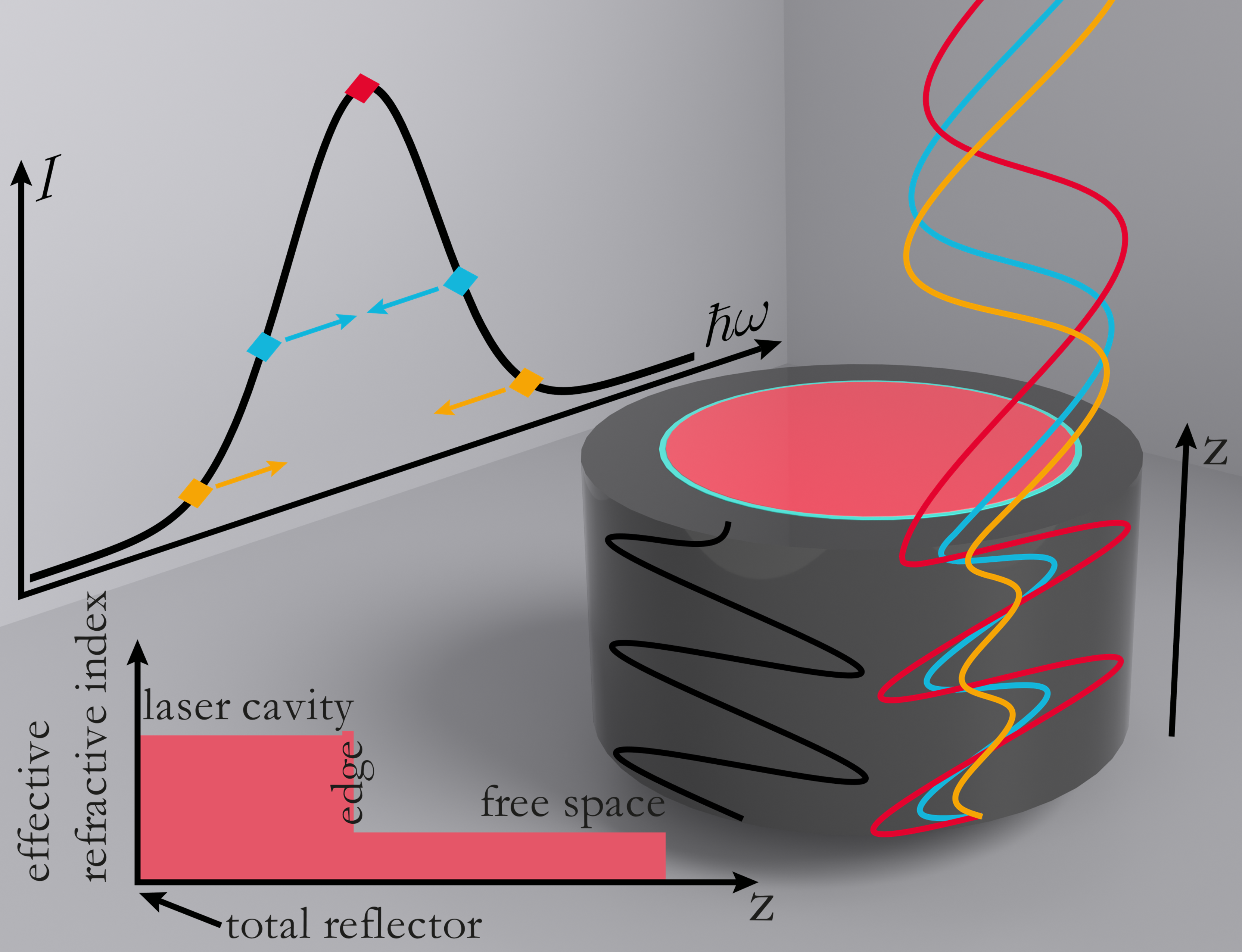}
	\caption{On the right is an illustration of an open cavity multimode model in comparison to the quasimode approach. The quasimode, shown in black,
	is, in good approximation, a composition of a continuum of many open cavity modes populating the combined system of resonator and free space, here
	exemplarily shown in red, blue and orange. The top left plot shows a qualitative dependence of the individual modes' intensities as a function of
	the eigenfrequencies of the "cold resonator" modes. The arrows illustrate the shift of the individual emission frequencies due to the partial mode
	locking. The bottom left plot highlights the effective refractive index for individual sections of the open cavity system used to model the
	resonator for the composite ansatz.}
	 \label{fig:Multimodes}
\end{figure}

\begin{figure} [!t]
	\centering
	\includegraphics[width=1.0\linewidth]{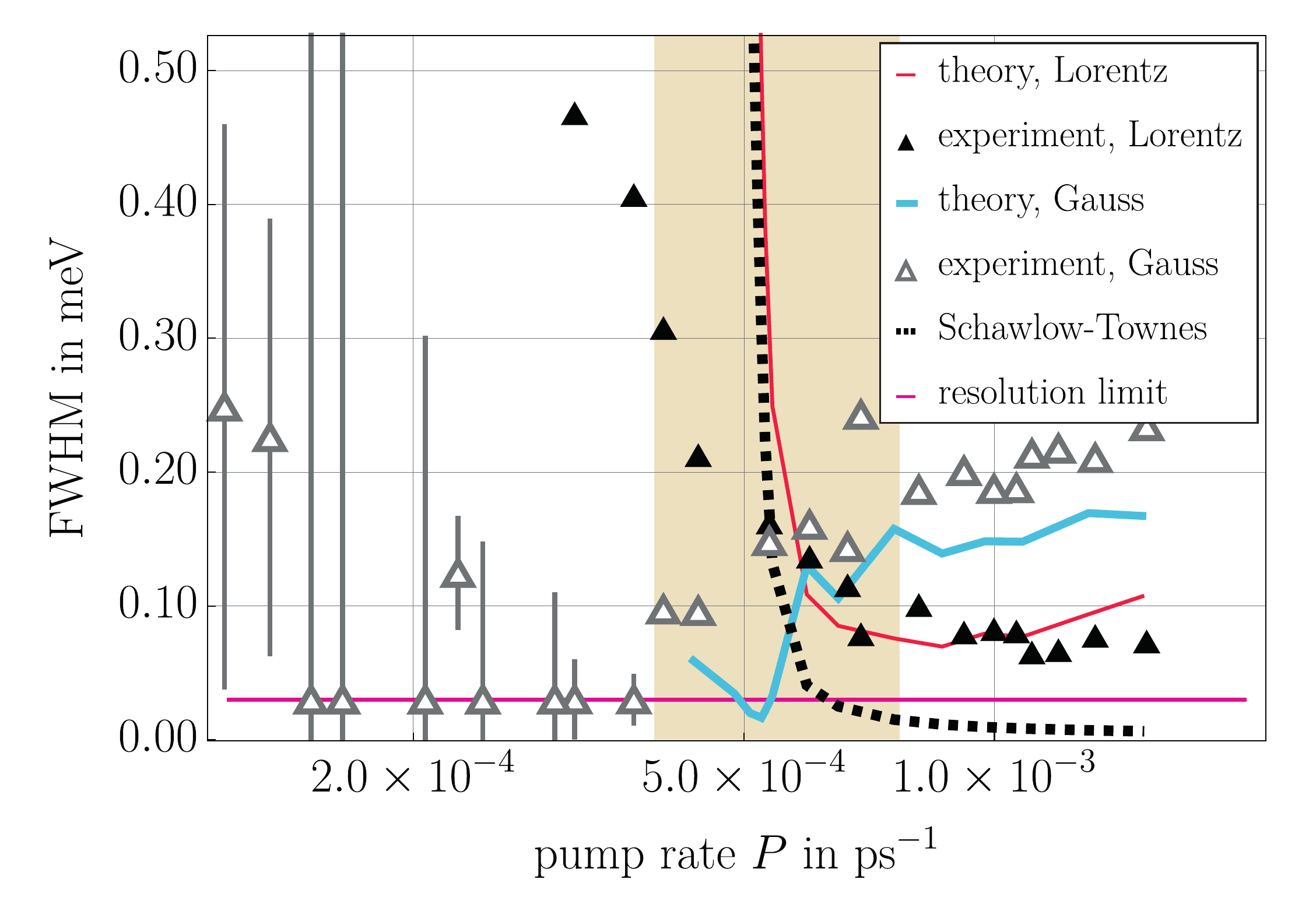}
	\caption{Comparison of the Lorentzian and Gaussian components extracted from the experimentally measured spectra and those calculated from the
	semiclassical multimode model. The multimode model clearly reproduces the transition to a Gaussian lineshape at the laser threshold in very good
	agreement with the experiment. Above the threshold, the FWHM is also correctly described. The linewidths stay above the resolution limit (0.03$\,$meV) of the $\mu$PL setup.}
	\label{fig:VoigtWeng}
\end{figure}

To provide an explanation for the observed lineshape change, we employed a mathematically more rigorous description of eigenmodes of an optical cavity with outcoupling losses. Here, the quasi-(Fox-Li) mode typically used in laser theory \cite{Sargent1974} is represented by a set of eigenmodes to the homogeneous wave equation that extend from inside the cavity out to free space, as illustrated in Fig.~\ref{fig:Multimodes}. As a result, a multimode laser theory is used to account for both linear and nonlinear contributions from the active medium. This is in contrast to the often used single-mode cQED treatment.  Here, the lineshape transition taking place at the lasing threshold arises from the intrinsic mechanisms of gain clamping and frequency locking of the combined system of the laser cavity and free space. For the evaluation of the theory, we derive equations of motion for the intensity and phase of the individual modes and the charge carrier density, giving a coupled system of equations that we solve numerically. We analyze the numerically obtained spectra (shown in Fig.~S3 in the SI), again using a Voigt profile and show the Lorentzian and Gaussian linewidth contribution together with those obtained from the measured spectra in Fig. \ref{fig:VoigtWeng}. Indeed, the spectra we obtain from the multimode approach reproduce the transition to a Gaussian component (gray curve) at the laser threshold in excellent agreement with the experimental data. The partial mode locking of the combined laser cavity and free space modes is the underlying mechanism that gives rise to the emergence of Gaussian lineshape component, pulling the lasing frequencies of the individual modes to the central emission frequency as schematically illustrated in Fig.~\ref{fig:Multimodes}.
As a result, a dominant Gaussian shape emerges \cite{Chow2021}. It should be noted that the appearance of the Gaussian component is indirectly related to the onset of the stimulated emission, as the mode locking increases in strength with rising output intensity.

While the semiclassical multimode ansatz successfully explains the line shape anomaly, it fails to describe the FWHM of the Lorentzian component of the spectrum below the laser threshold. This is a consequence of the way that spontaneous emission is included in a semiclassical theory. As such, it reflects a difference to the quantum-optical model, which contains spontaneous emission naturally due to the quantization of the light field. The semiclassical model shows, in contrast to the quantum optical model, the well known Schawlow-Townes behavior, as indicated by the black line in Fig.~\ref{fig:VoigtWeng}.

\section{Conclusions}

In summary, we provide new insight into the exciting physics of ultra-small semiconductor lasers with high $\beta$ factors. From a combined experimental and theoretical quantum-optical study, we have established high-$\beta$ lasing operation at telecom wavelength in a metallic nanocavity laser. Combined signatures in the autocorrelation function and the coherence time serve as a clear indicator for the laser transition. Furthermore, we report a lineshape anomaly in the emission spectra with a transition from a Lorentzian to a Gaussian shape at the laser threshold, directly reflected in the shape factor of the line profile. We have carefully ruled out extrinsic effects related to noise and inhomogeneous effects. Instead, we find that the Gaussian lineshape arises intrinsically from pulling of partly-locked modes. This insight is obtained from a complementary semiclassical theoretical approach that sacrifices the generally used closed-cavity quasi-mode, but employs a composite ansatz for the laser and free-space modes. Since the effect is related to the onset of stimulated emission, in principle, this makes it possible to identify the threshold in high-$\beta$ nanolasers solely from the emission spectra without the need of measuring the second-order photon-autocorrelation function. Limitations of the semiclassical open-cavity ansatz show in manifest in the description of the linewidth behavior in the low-excitation regime, which is correctly reproduced by our quantum-optical theory. This serves as a strong motivation for the development of new laser models that combine the best of both approaches, i.e. the quantum-optical treatment of the light field with the multi-mode description of cavity and free-space.

\section{Methods}

\subsection{Sample Fabrication}

First, an InGaAsP-wafer containing multiple (6) quantum wells is grown on an InP-substrate via metal organic chemical vapor deposition.
Electron beam lithography and dry etching techniques using a SiO${_2}$ hard mask are employed to form cylindrical pillars of the now
free standing gain material on the wafer. A 10$\,$nm-thin dielectric layer of Al$_{2}$O$_{3}$ is then deposited on top of the pillars,
followed by the capping with 100$\,$nm of silver to construct the metallic cavity. The deposition of the dielectric layer is an important step
in this procedure since it leads to the reduction of the significant optical losses in metals at visible and NIR wavelengths through dissipation
and surface carrier recombination. The last steps of the fabrication include gluing the Ag-coated cavity onto a silicon wafer, removing the InP
base and flipping the wafer $180^\circ$ to form the final nanolaser design. The whole procedure is repeated to fabricate 15x15 arrays of
nanolasers with diameters varying from 100 to 800$\,$nm.

\subsection{Experimental Configuration}

The nanolaser sample is mounted into a He-flow cryostat to enable low temperature operation with precise temperature control. Optical excitation is applied through a diode laser emitting at 785$\,$nm and operated at continuous wave (CW) mode. The excitation and collection of PL emission is
realized through a confocal arrangement with a microscope objective of NA 0.4 and a focal length of 10$\,$mm. In the detection path, the diode
laser light is blocked through a longpass filter allowing optical transmission at wavelengths above 1200$\,$nm. PL of the nanolaser is led
through a half-wavelength $(\lambda/2)$ plate (HWP) and a linear polarizer (lin. pol.) into a Czerny-Turner monochromator that, depending on the choice
of the grating (900$\,$grooves/mm at the finest grating), can reach a spectral resolution as good as 0.05$\,$nm in the first refractive order using
a cooled InGaAs 1D-array as a detector. Alternatively, the light emission can be directed to the fiber-based HBT configuration attached
to the exit slit of the monochromator, which allows us to investigate the statistical light properties by means of photon-autocorrelation
measurements. For this purpose, we employ two superconducting nanowire single photon detectors (SNSPDs) with a combined HBT resolution of 80$\,$ps. It should be noted that the 900$\,$grooves/mm prism set in the first refractive order leads to spectral filtering with an estimated window
of about 22$\,$pm at the fiber facet. Due to the much wider emission linewidth of the investigated nanolaser on the order of 0.4 to 2.5$\,$nm,
the $g^{(2)}$-measurements would suffer in that configuration from strong intensity fluctuations leading to artificial enhanced bunching in the
recorded data \cite{Kreinberg2020}. To overcome this problem, the groove prism is set to zero-order, basically operating as a mirror without
spectral resolution. An additional bandpass filter with a spectral window of 10$\,$nm centered at 1450$\,$nm is introduced into the detection
beam path. By tilting the filter with respect to the normal incidence, the central wavelength of the filter can be shifted to the appropriate
central wavelength (CWL) of the signal to ensure that only photons of a single lasing mode contribute to the correlations in the registered
data. All measurements in this study have been performed on a selected nanolaser with a diameter of 700$\,$nm.

\section*{Acknowledgements}

We thank the DFG for financial support via the projects Re2974/21-1 and Ja619/18-1. F.L. acknowledges funding from the University of Bremen Central Research Developing Fund (CRDF). The Tsinghua group acknowledges financial support from Beijing Innovation Centre for Future Chips at Tsinghua University, NSFC No 91750206 and No 61861136006. We further acknowledge technical support by the group of Tobias Heindel funded via the BMBF-project ”QuSecure” (Grant No. 13N14876) within the funding program Photonic Research Germany.

\section*{Disclosures} The authors declare no conflicts of interest.

\bibliographystyle{hieeetr}

\begin{thebibliography}{10}

\bibitem{Miller2009}
D.~A. Miller, ``Device requirements for optical interconnects to silicon
  chips,'' {\em Proceedings of the IEEE}, vol.~97, no.~7, pp.~1166--1185, 2009.

\bibitem{Ning2019}
C.-Z. Ning, ``{Semiconductor nanolasers and the size-energy-efficiency
  challenge: a review},'' {\em Advanced Photonics}, vol.~1, no.~1, pp.~1 -- 10,
  2019.

\bibitem{Rodt2021}
S.~Rodt and S.~Reitzenstein, ``Integrated nanophotonics for the development of
  fully functional quantum circuits based on on-demand single-photon
  emitters,'' {\em {APL} Photonics}, vol.~6, p.~010901, Jan. 2021.

\bibitem{Wang2018}
K.~Wang, S.~Wang, S.~Xiao, and Q.~Song, ``Recent advances in perovskite
  micro-and nanolasers,'' {\em Advanced Optical Materials}, vol.~6, no.~18,
  p.~1800278, 2018.

\bibitem{Jeong2020}
K.-Y. Jeong, M.-S. Hwang, J.~Kim, J.-S. Park, J.~M. Lee, and H.-G. Park,
  ``Recent progress in nanolaser technology,'' {\em Advanced Materials},
  vol.~32, no.~51, p.~2001996, 2020.

\bibitem{Liang2020}
Y.~Liang, C.~Li, Y.-Z. Huang, and Q.~Zhang, ``Plasmonic nanolasers in on-chip
  light sources: Prospects and challenges,'' {\em ACS nano}, vol.~14, no.~11,
  pp.~14375--14390, 2020.

\bibitem{Azzam2020}
S.~I. Azzam, A.~V. Kildishev, R.-M. Ma, C.-Z. Ning, R.~Oulton, V.~M. Shalaev,
  M.~I. Stockman, J.-L. Xu, and X.~Zhang, ``Ten years of spasers and plasmonic
  nanolasers,'' {\em Light: Science \& Applications}, vol.~9, no.~1, pp.~1--21,
  2020.

\bibitem{Deng2021}
H.~Deng, G.~L. Lippi, J.~M{\o}rk, J.~Wiersig, and S.~Reitzenstein, ``Physics
  and applications of high-$\beta$ micro-and nanolasers,'' {\em Advanced
  Optical Materials}, vol.~9, no.~19, p.~2100415, 2021.

\bibitem{Bjork1994}
G.~Bj{\"o}rk, A.~Karlsson, and Y.~Yamamoto, ``Definition of a laser
  threshold,'' {\em Physical Review A}, vol.~50, no.~2, p.~1675, 1994.

\bibitem{Ning2013}
C.-Z. Ning, ``What is laser threshold?,'' {\em IEEE Journal of Selected Topics
  in Quantum Electronics}, vol.~19, no.~4, pp.~1503604--1503604, 2013.

\bibitem{Strauf2006}
S.~Strauf, K.~Hennessy, M.~Rakher, Y.-S. Choi, A.~Badolato, L.~Andreani, E.~Hu,
  P.~Petroff, and D.~Bouwmeester, ``Self-tuned quantum dot gain in photonic
  crystal lasers,'' {\em Physical review letters}, vol.~96, no.~12, p.~127404,
  2006.

\bibitem{Samuel2009}
I.~D. Samuel, E.~B. Namdas, and G.~A. Turnbull, ``How to recognize lasing,''
  {\em Nature Photonics}, vol.~3, no.~10, pp.~546--549, 2009.

\bibitem{Chow2014}
W.~W. Chow, F.~Jahnke, and C.~Gies, ``Emission properties of nanolasers during
  the transition to lasing,'' {\em Light: Science \& Applications}, vol.~3,
  no.~8, pp.~e201--e201, 2014.

\bibitem{Kreinberg2017}
S.~Kreinberg, W.~W. Chow, J.~Wolters, C.~Schneider, C.~Gies, F.~Jahnke,
  S.~H{\"o}fling, M.~Kamp, and S.~Reitzenstein, ``Emission from quantum-dot
  high-$\beta$ microcavities: transition from spontaneous emission to lasing
  and the effects of superradiant emitter coupling,'' {\em Light: Science \&
  Applications}, vol.~6, no.~8, pp.~e17030--e17030, 2017.

\bibitem{Jagsch2018}
S.~T. Jagsch, N.~V. Trivi{\~n}o, F.~Lohof, G.~Callsen, S.~Kalinowski, I.~M.
  Rousseau, R.~Barzel, J.-F. Carlin, F.~Jahnke, R.~Butt{\'e}, {\em et~al.}, ``A
  quantum optical study of thresholdless lasing features in high-$\beta$
  nitride nanobeam cavities,'' {\em Nature communications}, vol.~9, no.~1,
  pp.~1--7, 2018.

\bibitem{Reeves2018}
L.~Reeves, Y.~Wang, and T.~F. Krauss, ``2d material microcavity light emitters:
  to lase or not to lase?,'' {\em Advanced Optical Materials}, vol.~6, no.~19,
  p.~1800272, 2018.

\bibitem{Wu2019}
H.~Wu, Y.~Gao, P.~Xu, X.~Guo, P.~Wang, D.~Dai, and L.~Tong, ``Plasmonic
  nanolasers: Pursuing extreme lasing conditions on nanoscale,'' {\em Advanced
  Optical Materials}, vol.~7, no.~17, p.~1900334, 2019.

\bibitem{Chow2018}
W.~W. Chow and S.~Reitzenstein, ``Quantum-optical influences in
  optoelectronics—an introduction,'' {\em Applied Physics Reviews}, vol.~5,
  no.~4, p.~041302, 2018.

\bibitem{Schawlow1958}
A.~L. Schawlow and C.~H. Townes, ``Infrared and optical masers,'' {\em Physical
  Review}, vol.~112, no.~6, p.~1940, 1958.

\bibitem{Scully1966}
M.~Scully and W.~Lamb~Jr, ``Quantum theory of an optical maser,'' {\em Physical
  Review Letters}, vol.~16, no.~19, p.~853, 1966.

\bibitem{Haken1966}
H.~Haken, ``Theory of intensity and phase fluctuations of a homogeneously
  broadened laser,'' {\em Zeitschrift f{\"u}r Physik}, vol.~190, no.~3,
  pp.~327--356, 1966.

\bibitem{Lax1966}
M.~Lax, P.~Kelley, and P.~Tannenwald, ``Physics of quantum electronics,'' {\em
  New York}, p.~735, 1966.

\bibitem{Haug1967}
H.~Haug and H.~Haken, ``Theory of noise in semiconductor laser emission,'' {\em
  Zeitschrift f{\"u}r Physik A Hadrons and nuclei}, vol.~204, no.~3,
  pp.~262--275, 1967.

\bibitem{Glauser2014}
M.~Glauser, C.~Mounir, G.~Rossbach, E.~Feltin, J.-F. Carlin, R.~Butt{\'e}, and
  N.~Grandjean, ``Ingan/gan quantum wells for polariton laser diodes: Role of
  inhomogeneous broadening,'' {\em Journal of Applied Physics}, vol.~115,
  no.~23, p.~233511, 2014.

\bibitem{Stephan2005}
G.~St{\'e}phan, T.~Tam, S.~Blin, P.~Besnard, and M.~T{\^e}tu, ``Laser line
  shape and spectral density of frequency noise,'' {\em Physical Review A},
  vol.~71, no.~4, p.~043809, 2005.

\bibitem{Septon2019}
T.~Septon, A.~Becker, S.~Gosh, G.~Shtendel, V.~Sichkovskyi, F.~Schnabel,
  A.~Seng{\"u}l, M.~Bjelica, B.~Witzigmann, J.~P. Reithmaier, {\em et~al.},
  ``Large linewidth reduction in semiconductor lasers based on atom-like gain
  material,'' {\em Optica}, vol.~6, no.~8, pp.~1071--1077, 2019.

\bibitem{Hill2009}
M.~T. Hill, M.~Marell, E.~S. Leong, B.~Smalbrugge, Y.~Zhu, M.~Sun, P.~J.
  Van~Veldhoven, E.~J. Geluk, F.~Karouta, Y.-S. Oei, {\em et~al.}, ``Lasing in
  metal-insulator-metal sub-wavelength plasmonic waveguides,'' {\em Optics
  express}, vol.~17, no.~13, pp.~11107--11112, 2009.

\bibitem{Oulton2009}
R.~F. Oulton, V.~J. Sorger, T.~Zentgraf, R.-M. Ma, C.~Gladden, L.~Dai,
  G.~Bartal, and X.~Zhang, ``Plasmon lasers at deep subwavelength scale,'' {\em
  Nature}, vol.~461, no.~7264, pp.~629--632, 2009.

\bibitem{Noginov2009}
M.~Noginov, G.~Zhu, A.~Belgrave, R.~Bakker, V.~Shalaev, E.~Narimanov, S.~Stout,
  E.~Herz, T.~Suteewong, and U.~Wiesner, ``Demonstration of a spaser-based
  nanolaser,'' {\em Nature}, vol.~460, no.~7259, pp.~1110--1112, 2009.

\bibitem{Khajavikhan2012}
M.~Khajavikhan, A.~Simic, M.~Katz, J.~Lee, B.~Slutsky, A.~Mizrahi, V.~Lomakin,
  and Y.~Fainman, ``Thresholdless nanoscale coaxial lasers,'' {\em Nature},
  vol.~482, no.~7384, pp.~204--207, 2012.

\bibitem{Kwon2012}
S.-H. Kwon, ``Deep subwavelength plasmonic whispering-gallery-mode cavity,''
  {\em Optics express}, vol.~20, no.~22, pp.~24918--24924, 2012.

\bibitem{Brown1956}
R.~H. Brown and R.~Q. Twiss, ``Correlation between photons in two coherent
  beams of light,'' {\em Nature}, vol.~177, no.~4497, pp.~27--29, 1956.

\bibitem{Kreinberg2020}
S.~Kreinberg, K.~Laiho, F.~Lohof, W.~E. Hayenga, P.~Holewa, C.~Gies,
  M.~Khajavikhan, and S.~Reitzenstein, ``Thresholdless transition to coherent
  emission at telecom wavelengths from coaxial nanolasers with excitation power
  dependent $\beta$-factors,'' {\em Laser \& Photonics Reviews}, vol.~14,
  no.~12, p.~2000065, 2020.

\bibitem{Ulrich2007}
S.~Ulrich, C.~Gies, S.~Ates, J.~Wiersig, S.~Reitzenstein, C.~Hofmann,
  A.~L{\"o}ffler, A.~Forchel, F.~Jahnke, and P.~Michler, ``Photon statistics of
  semiconductor microcavity lasers,'' {\em Physical review letters}, vol.~98,
  no.~4, p.~043906, 2007.

\bibitem{Wiersig2009}
J.~Wiersig, C.~Gies, F.~Jahnke, M.~A{\ss}mann, T.~Berstermann, M.~Bayer,
  C.~Kistner, S.~Reitzenstein, C.~Schneider, S.~H{\"o}fling, {\em et~al.},
  ``Direct observation of correlations between individual photon emission
  events of a microcavity laser,'' {\em Nature}, vol.~460, no.~7252,
  pp.~245--249, 2009.

\bibitem{Hostein2010}
R.~Hostein, R.~Braive, L.~Le~Gratiet, A.~Talneau, G.~Beaudoin,
  I.~Robert-Philip, I.~Sagnes, and A.~Beveratos, ``Demonstration of coherent
  emission from high-$\beta$ photonic crystal nanolasers at room temperature,''
  {\em Optics letters}, vol.~35, no.~8, pp.~1154--1156, 2010.

\bibitem{Assmann2010}
M.~A{\ss}mann, F.~Veit, M.~Bayer, C.~Gies, F.~Jahnke, S.~Reitzenstein,
  S.~H{\"o}fling, L.~Worschech, and A.~Forchel, ``Ultrafast tracking of
  second-order photon correlations in the emission of quantum-dot
  microresonator lasers,'' {\em Physical Review B}, vol.~81, no.~16, p.~165314,
  2010.

\bibitem{Hopfmann2013}
C.~Hopfmann, F.~Albert, C.~Schneider, S.~H{\"o}fling, M.~Kamp, A.~Forchel,
  I.~Kanter, and S.~Reitzenstein, ``Nonlinear emission characteristics of
  quantum dot--micropillar lasers in the presence of polarized optical
  feedback,'' {\em New Journal of Physics}, vol.~15, no.~2, p.~025030, 2013.

\bibitem{Takiguchi2016}
M.~Takiguchi, H.~Taniyama, H.~Sumikura, M.~D. Birowosuto, E.~Kuramochi,
  A.~Shinya, T.~Sato, K.~Takeda, S.~Matsuo, and M.~Notomi, ``Systematic study
  of thresholdless oscillation in high-$\beta$ buried multiple-quantum-well
  photonic crystal nanocavity lasers,'' {\em Optics express}, vol.~24, no.~4,
  pp.~3441--3450, 2016.

\bibitem{Ota2017}
Y.~Ota, M.~Kakuda, K.~Watanabe, S.~Iwamoto, and Y.~Arakawa, ``Thresholdless
  quantum dot nanolaser,'' {\em Optics express}, vol.~25, no.~17,
  pp.~19981--19994, 2017.

\bibitem{Vyshnevyy2018}
A.~A. Vyshnevyy and D.~Y. Fedyanin, ``Lasing threshold of thresholdless and
  non-thresholdless metal-semiconductor nanolasers,'' {\em Optics express},
  vol.~26, no.~25, pp.~33473--33483, 2018.

\bibitem{Gies2019}
C.~Gies and S.~Reitzenstein, ``Quantum dot micropillar lasers,'' {\em
  Semiconductor Science and Technology}, vol.~34, no.~7, p.~073001, 2019.

\bibitem{Reitzenstein2008a}
S.~Reitzenstein, T.~Heindel, C.~Kistner, A.~Rahimi-Iman, C.~Schneider,
  S.~H{\"o}fling, and A.~Forchel, ``Low threshold electrically pumped quantum
  dot-micropillar lasers,'' {\em Applied Physics Letters}, vol.~93, no.~6,
  p.~061104, 2008.

\bibitem{Redlich2016}
C.~Redlich, B.~Lingnau, S.~Holzinger, E.~Schlottmann, S.~Kreinberg,
  C.~Schneider, M.~Kamp, S.~H{\"o}fling, J.~Wolters, S.~Reitzenstein, {\em
  et~al.}, ``Mode-switching induced super-thermal bunching in quantum-dot
  microlasers,'' {\em New Journal of Physics}, vol.~18, no.~6, p.~063011, 2016.

\bibitem{Sargent1974}
M.~Sargent~III, M.~Scully, and W.~Lamb~Jr, ``Laser physics addison-wesley,''
  {\em Reading, Mass}, 1974.

\bibitem{Chow2021}
W.~W. Chow, Y.~Wan, J.~E. Bowers, and F.~Grillot, ``Analysis of the spontaneous
  emission limited linewidth of an integrated iii-v/sin laser,'' 2021,
  2112.11403 (in revision).

\end{thebibliography}

\end{document}


\maketitle

\newpage


\section*{A quantum optical nanolaser model with quantum wells as active material}

\section{System Hamiltonian and Heisenberg Equation of Motion}

The theory is built on an equation of motion approach for the expectation values of observables described by photon and carrier operators using the
system's Hamiltonian in conjunction with Lindblad terms to model resonator and cavity decay losses.
The Hamilton operator considers three contributions:

\begin{align*}
	\hat{\mathcal{H}}_{\text{car.}}
	&= \sum_{\vec{k},s} \left( \eps{k}{s}{c} \cc{k}{s} \ca{k}{s} + \eps{k}{s}{v} \vc{k}{s} \va{k}{s} \right)
\quad, \\
	\hat{\mathcal{H}}_{\text{pho.}}
	&= \sum_{\xi} \hbar \omega_{\xi} \left( \bc{\xi} \ba{\xi} + \frac{1}{2} \right)
\quad, \\
	\hat{\mathcal{H}}_{\text{int.}}
	&= i \hbar \sum_{\vec{k},s,\xi,j}
	\left(
	\g{k}{s}{\xi,j} \bc{\xi} \vc{k}{s} \ca{k}{s} - \gast{k}{s}{\xi,j} \ba{\xi} \cc{k}{s} \va{k}{s}
	\right)
\quad, \\
	\text{with} ~~~ \g{k}{s}{\xi,j} &= \frac{g_{0,j}}{1 + \frac{\hbar^2 |\vec{k}|^2}{2 E_{G}}
	\left( \frac{1}{m_{e}} + \frac{1}{m_{h}} \right)} \quad.
\end{align*}

Here, $\vec{k},s$ are the indices denoting momentum and spin of the carriers and $\xi$ is the quantum number characterising the
photons of an individual mode. $\bc{\xi}$ and $\ba{\xi}$ are the photonic creation and annihilation operators; $\vc{k}{s}$ / $\cc{k}{s}$
and $\va{k}{s}$ / $\ca{k}{s}$ are the carrier creation and annihilation operators in the valence and conduction band, respectively,
which obey the following commutator relations:

\begin{align*}
	\left[ \bc{\xi} , \bc{\xi'} \right]
	= \left[ \ba{\xi} , \ba{\xi'} \right] = 0
	&~~~ , ~~~ \left[ \ba{\xi} , \bc{\xi'} \right] = \delta_{\vphantom{\vec{k}} \xi,\xi'}
\quad, \\
	\left\{ \cc{k}{s} , \cc{k'}{s'} \right\}
	= \left\{ \ca{k}{s} , \ca{k'}{s'} \right\} = 0
	&~~~ , ~~~ \left\{ \ca{k}{s} , \cc{k'}{s'} \right\} = \delta_{\vec{k},\vec{k'}}
	\delta_{\vphantom{\vec{k}} s,s'}
\quad, \\
	\left\{ \cc{k}{s} , \vc{k'}{s'} \right\}
	= \left\{ \ca{k}{s} , \va{k'}{s'} \right\} = 0
	&~~~ , ~~~ \left\{ \ca{k}{s} , \vc{k'}{s'} \right\} = \delta_{\vec{k},\vec{k'}}
	\delta_{\vphantom{\vec{k}} s,s'} \delta_{\vphantom{\vec{k}} c,v} \quad.
\end{align*}

The additional index $j$ denotes individual quantum wells to consider the possibility of their coupling to the lasing mode with different
light-matter interaction-strenghts. In order to simplify the notation, the derivations in the following sections
will only consider the same interaction-strengths for each of the quantum wells, allowing us to drop the $j$-index and the associated sum.

\newpage

For the energies of electrons and holes, a parabolic two-band model is used:

\begin{align*}
	\eps{k}{s}{c} &= \eps{k}{s}{e} + E_{G} = \frac{\hbar^2 |\vec{k}|^2}{2 m_{e}} + E_{G}
\quad, \\
	\eps{k}{s}{v} &= - \eps{k}{s}{h} = - \frac{\hbar^2 |\vec{k}|^2}{2 m_{h}}
\quad, \\
	\eps{k}{s}{c} - \eps{k}{s}{v} - \hbar \omega_{\xi_L} &= \frac{\hbar^2 |\vec{k}|^2}{2} \left( \frac{1}{m_e} + \frac{1}{m_h} \right)
	- \hbar \delta \quad.
\end{align*}

\begin{table}[ht!]
\begin{center}
\begin{tabular}{ccc} \toprule
    {parameter} 		& {description} 						& {assigned value} 										\\ \toprule
    {$m_{0}$}			& {free electron mass}					& {$5.686 \cdot 10^{-3}\,$meV ps$^2$ nm$^{-2}$} 		\\
    {$m_{e}$}  			& {electron mass} 						& {$0.057\,m_{0}$} 										\\
    {$m_{h}$}  			& {hole mass}  							& {$0.170\,m_{0}$}  									\\
    {$E_{G}$}  			& {band gap energy}  					& {$816\,$meV} 											\\ \midrule
    {$k_{max}$}			& {maximum k-value}						& {$2.0\,$nm$^{-1}$}									\\ \midrule
    {$\hbar \delta  = \hbar \omega_{\xi_{L}} - E_{G}$}  		& {detuning}  				& {$10\,$meV} 				\\ \bottomrule
\end{tabular}
\end{center}
\caption*{Tab. S1: Relevant bandstructure parameters utilised for the semiconductor laser theory described here.}
\end{table}

\begin{figure}[ht!]
\begin{center}
{\fbox{\includegraphics[width=0.70\textwidth]{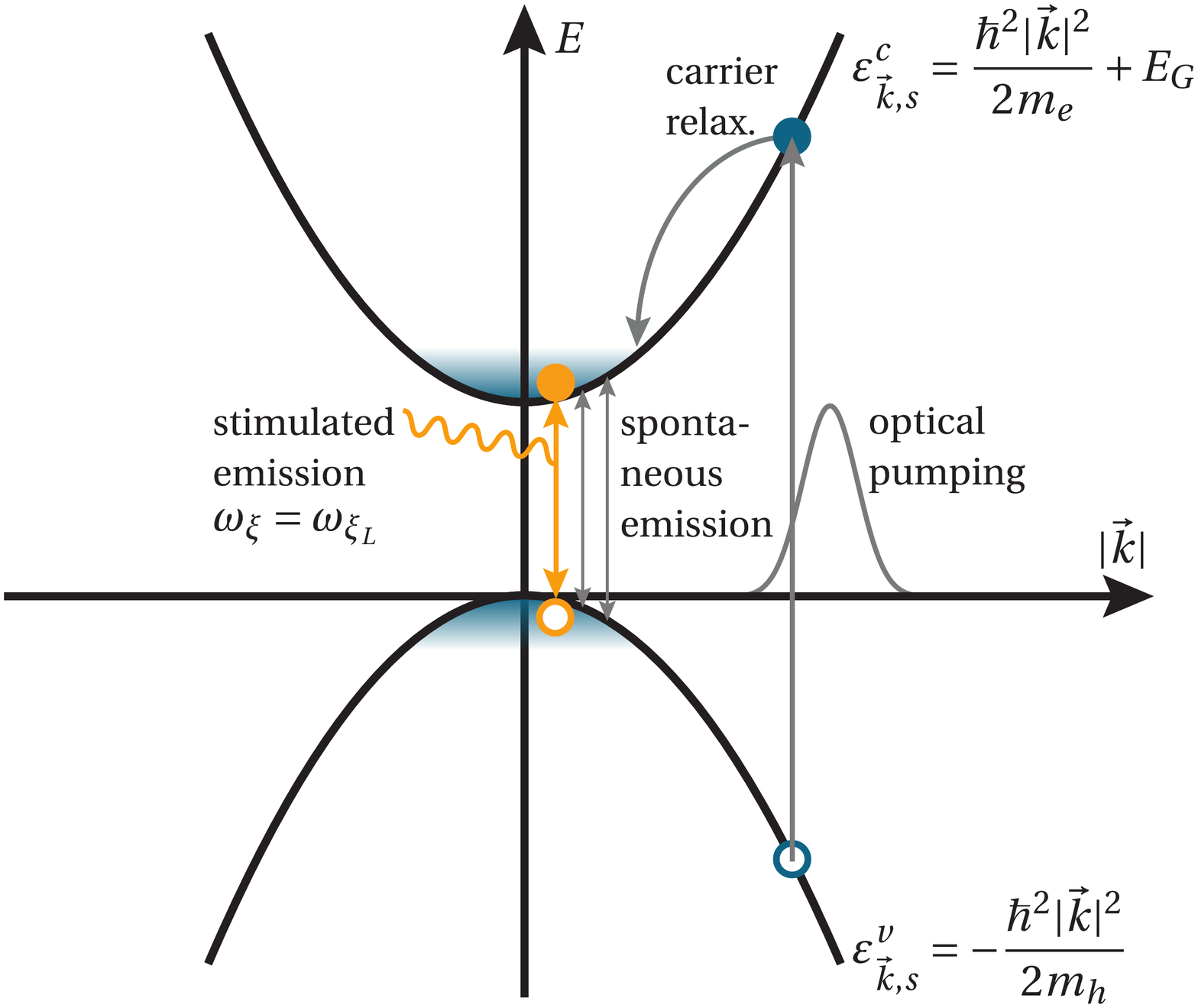}}}
\end{center}
\caption*{Fig. S1: Band structure and processes included in the quantum optical nanolaser theory.}
\end{figure}

In order to derive equations that describe the temporal dynamics of physical observables, such as the photon number, depending on the
rate at which the system is pumped, the Heisenberg equation of motion (EoM) with Ehrenfest's theorem for the calculation of the
expectation values of quantum mechanical operators $\hat{A}$ is employed:

\newpage

\begin{align*}
	\frac{\text{d}}{\text{d} t} \langle \hat{A} \rangle = \frac{i}{\hbar}
	\langle \left[
	\hat{\mathcal{H}}_{\text{car.}} + \hat{\mathcal{H}}_{\text{pho.}} + \hat{\mathcal{H}}_{\text{int.}}, \hat{A}
	\rangle \right] \quad.
\end{align*}

The full quantisation of the light field will give access to higher-order correlations, but considering light-matter-interaction terms also
gives rise to a hierarchy problem that requires a truncation of the resulting equations at a specific level.
Depending on the number of particles involved in these correlations this is either the doublet-level (two particles) or the
quadruplet-level (four particles). Note that due to the necessity of particle number conservation, we use the convention that either
two fermionic operators or one bosonic operator account for a single particle.

\section{Photon number $n_{\xi_{L}}$ and doublet-level laser equations}

Here, Lindblad terms have been considered for the modelling of resonator and cavity decay losses with the only generator of interest being
$\hat{V}_{}^{\dagger} = \sqrt{2 \tilde{\kappa}_{\xi}} \bc{\xi}$ using $\tilde{\kappa}_{\xi} / \hbar = \kappa_{\xi}$, and

\begin{align*}
	\frac{\text{d}}{\text{d}t} \langle \hat{A} \rangle
	&= \frac{i}{\hbar}
	\langle \left[ \hat{\mathcal{H}}_\text{car.} + \hat{\mathcal{H}}_\text{pho.} + \hat{\mathcal{H}}_\text{int.},\hat{A} \right] \rangle
	+ \frac{1}{2 \hbar} \hat{V}_{}^{\dagger} \left( [ \hat{A},\hat{V}_{}^{} ]
	+ [ \hat{V}_{}^{\dagger}, \hat{A} ] \hat{V}_{}^{} \right)
\\
	&= \frac{i}{\hbar}
	\langle \left[ \hat{\mathcal{H}}_\text{car.} + \hat{\mathcal{H}}_\text{pho.} + \hat{\mathcal{H}}_\text{int.},\hat{A} \right] \rangle
	+ \hat{V}_{}^{\dagger} \hat{A} \hat{V}_{}^{} - \frac{1}{2} \left\{ \hat{V}_{}^{\dagger} \hat{V}_{}^{}, \hat{A} \right\} \quad.
\end{align*}

Application of this equation to the photon number $n_{\xi} = \langle \bc{\xi} \ba{\xi} \rangle$, the electron and hole populations
$f_{\vec{k},s}^{e} = \langle \cc{k}{s} \ca{k}{s} \rangle$, $f_{\vec{k},s}^{h} = 1 - \langle \vc{k}{s} \va{k}{s} \rangle$,
as well as the photon-assisted polarisation $\psi_{\vec{k},s,\xi}^{0} = \langle \bc{\xi} \vc{k}{s} \ca{k}{s} \rangle$ operators
leads to the first set of coupled doublet-level laser equations as follows:

\begin{align*}
	\frac{\text{d}}{\text{d}t} \left( \langle \cc{k}{s} \ca{k}{s} \rangle \right)
	=
	&- 2 \sum_{\xi'} \text{Re} \left[ \g{k}{s}{\xi'}
	\langle \bc{\xi'} \vc{k}{s} \ca{k}{s} \rangle \right]
\quad, \\
	\frac{\text{d}}{\text{d}t} \left( 1 - \langle \vc{k}{s} \va{k}{s} \rangle \right)
	=
	&- 2 \sum_{\xi'} \text{Re} \left[ \g{k}{s}{\xi'}
	\langle \bc{\xi'} \vc{k}{s} \ca{k}{s} \rangle \right]
\quad, \\
	\frac{\text{d}}{\text{d}t} \left( \langle \bc{\xi} \ba{\xi} \rangle \right)
	=
	&+ 2 \sum_{\vec{k'},s'} \text{Re} \left[ \g{k'}{s'}{\xi}
	\langle \bc{\xi} \vc{k'}{s'} \ca{k'}{s'} \rangle \right]
	- 2 \kappa_{\xi} \langle \bc{\xi} \ba{\xi} \rangle
\quad, \\
	\frac{\text{d}}{\text{d}t} \left( \langle \bc{\xi} \vc{k}{s} \ca{k}{s} \rangle \right)
	=
	&- \frac{i}{\hbar} \left( \eps{k}{s}{c} - \eps{k}{s}{v} - \hbar \omega_{\xi} \right)
	\langle \bc{\xi} \vc{k}{s} \ca{k}{s} \rangle
	+ \sum_{\xi'} \gast{k}{s}{\xi'}
	\left(
	\langle \bc{\xi} \ba{\xi'} \cc{k}{s} \ca{k}{s} \rangle
	- \langle \bc{\xi} \ba{\xi'} \vc{k}{s} \va{k}{s} \rangle
	\right)
	\\
	&- \sum_{\vec{k'},s',\xi'} \gast{k'}{s'}{\xi'} \langle \cc{k'}{s'} \vc{k}{s} \va{k'}{s'} \ca{k}{s} \rangle
	\delta_{\vphantom{\vec{k}} \xi,\xi'} - \kappa_{\xi} \langle \bc{\xi} \vc{k}{s} \ca{k}{s} \rangle \quad.
\end{align*}

Pumping of the system and thus driving of carriers is simulated using an incoherent Gaussian carrier distribution $ F_{\vec{k},s}^{p}$
created at high $|\vec{k}|$-values with an overall pump rate $P$. \\
The relaxation of carriers requires microscopic treatment to accurately account for phonon and carrier interactions.
However, since in this context only stationary solutions are of primary interest, we model a constant
relaxation rate $\gamma_{rel}$ which redistributes carriers from a non-equilibrium state towards a quasi-equilibrium state described via
the Fermi-Dirac distribution $F_{\vec{k},s}^{D}$.
In addition, $\Gamma$ is a phenomenological dephasing term introduced in agreement with \cite[eq. (2.25)]{chow1999semiconductor}. \\

\newpage

The photon-assisted polarisation couples to higher-order correlations via the terms
$\langle \bc{\xi} \ba{\xi'} \cc{k}{s} \ca{k}{s} \rangle$ and $\langle \bc{\xi} \ba{\xi'} \vc{k}{s} \va{k}{s} \rangle$.
According to the cluster expansion approach \cite[p. 313]{kira2011semiconductor} we can then factorise:

\begin{align*}
	\langle \bc{\xi} \ba{\xi'} \cc{k}{s} \ca{k}{s} \rangle
	&= \langle \bc{\xi} \ba{\xi'} \rangle \langle \cc{k}{s} \ca{k}{s} \rangle
	+ \delta \langle \bc{\xi} \ba{\xi'} \cc{k}{s} \ca{k}{s} \rangle
\quad, \\
	\langle \bc{\xi} \ba{\xi'} \vc{k}{s} \va{k}{s} \rangle
	&= \langle \bc{\xi} \ba{\xi'} \rangle \langle \vc{k}{s} \va{k}{s} \rangle
	+ \delta \langle \bc{\xi} \ba{\xi'} \vc{k}{s} \va{k}{s} \rangle \quad.
\end{align*}

In this model, the source term of spontaneous emission is the inter-band correlation \cite{baer2006luminescence,gies2007semiconductor}
with its associated factorisation:

\begin{align*}
	\langle \cc{k'}{s'} \vc{k}{s} \va{k'}{s'} \ca{k}{s} \rangle
	=
	\langle \cc{k'}{s'} \ca{k}{s} \rangle \langle \vc{k}{s} \va{k'}{s'} \rangle
	- \langle \cc{k'}{s'} \va{k'}{s'} \rangle \langle \vc{k}{s} \ca{k}{s} \rangle
	+ \delta \langle \cc{k'}{s'} \vc{k}{s} \va{k'}{s'} \ca{k}{s} \rangle \quad.
\end{align*}

Treating the light-matter-interaction quantum mechanically leads to a coupling of the conduction and valence band populations via the
photon-assisted polarisation. These coupled populations are only driven incoherently by an optical pump pulse at high $|\vec{k}|$-values.
Due to the absence of an external field, single-photon expectation values are set to zero:
$\langle \bc{\xi} \rangle = \langle \ba{\xi} \rangle = 0$. \cite{kira1998microscopic}
Since there is no electric field coherently driving polarisations (e.g. $\langle \vc{k}{s} \ca{k}{s} \rangle$) as a consequence of the pump
process, these are neglected here as well, which is also supported by the fact that carrier relaxation opposes the build-up of
inter-band polarisations. \\
Those expectation values of operators featuring two carriers in the same band (e.g. $\langle \cc{k'}{s'} \ca{k}{s} \rangle$) are
restricted to diagonal elements, allowing only populations to contribute \cite[eq. (72)]{kira1999quantum};
that is, because off-diagonal population-like quantities are not coupling to the conduction and valence band populations driving the
system's dynamics. \\
Furthermore, the influence of pure emitter-emitter correlations is neglected, leading us to drop
the $\delta$-term. For the final set of doublet-level laser equations this allows the transition:

\begin{align*}
	- \sum_{\vec{k'},s',\xi'} \gast{k'}{s'}{\xi'} \langle \cc{k'}{s'} \vc{k}{s} \va{k'}{s'} \ca{k}{s} \rangle
	\delta_{\vphantom{\vec{k}} \xi,\xi'}
	~~~ \rightarrow ~~~
	+ ~~ \gast{k}{s}{\xi} \langle \cc{k}{s} \ca{k}{s} \rangle \left( 1 - \langle \vc{k}{s} \va{k}{s} \rangle \right) \quad.
\end{align*}

In the quantum optical nanolaser model presented here, we consider the interaction of one lasing mode ($\xi = \xi_{L}$) with the gain material.
For all other emission channels, only spontaneous recombination is considered as these have no cavity feedback. As a consequence, the corresponding
photon numbers ($n_{\xi} = \langle \bc{\xi} \ba{\xi} \rangle, ~ \forall ~ \xi \neq \xi_{L}$) are negligible.
This allows to adiabatically eliminate $\langle \bc{\xi} \vc{k}{s} \ca{k}{s} \rangle$ for the non-lasing modes and to re-express the sum over
$\xi'$ occurring in the population equations by summarising all radiative losses via the rate:

\begin{align*}
	\gamma_{nl}
	= 2 \sum_{\xi' \neq \xi_{L}} \frac{\left| \g{k}{s}{\xi'} \right|^2 \left( \kappa_{\xi'} + \Gamma \right)}
	{\left( \kappa_{\xi'} + \Gamma \right)^2 + \hbar^{-2} \left( \eps{k}{s}{c} - \eps{k}{s}{v} - \hbar \omega_{\xi'} \right)}
\end{align*}

\newpage

Taking into consideration the outlined effects eventually leads to the final version of the Quantum Laser Equations on the
doublet-level that allow the calculation of the nanolaser's input-output characteristics: \\

\begin{align*}
	\frac{\text{d}}{\text{d}t} \left( \langle \cc{k}{s} \ca{k}{s} \rangle \right)
	=
	&- 2 \text{Re} \left[ \g{k}{s}{\xi_{L}}
	\langle \bc{\xi_{L}} \vc{k}{s} \ca{k}{s} \rangle \right]
	\\
	&- \gamma_{nl} \langle \cc{k}{s} \ca{k}{s} \rangle \left( 1 - \langle \vc{k}{s} \va{k}{s} \rangle \right)
	- \gamma_{rel} \left( \langle \cc{k}{s} \ca{k}{s} \rangle - F_{\vec{k},s}^{D} \right)
	+ P F_{\vec{k},s}^{p} \left( 1 - \langle \cc{k}{s} \ca{k}{s} \rangle \right)
\quad, \\
	\frac{\text{d}}{\text{d}t} \left( 1 - \langle \vc{k}{s} \va{k}{s} \rangle \right)
	=
	&- 2 \text{Re} \left[ \g{k}{s}{\xi_{L}}
	\langle \bc{\xi_{L}} \vc{k}{s} \ca{k}{s} \rangle \right]
	\\
	&- \gamma_{nl} \langle \cc{k}{s} \ca{k}{s} \rangle \left( 1 - \langle \vc{k}{s} \va{k}{s} \rangle \right)
	- \gamma_{rel} \left( 1 - \langle \vc{k}{s} \va{k}{s} \rangle - F_{\vec{k},s}^{D} \right)
	+ P F_{\vec{k},s}^{p} \left( \langle \vc{k}{s} \va{k}{s} \rangle \right)
\quad, \\
	\frac{\text{d}}{\text{d}t} \left( \langle \bc{\xi_{L}} \ba{\xi_{L}} \rangle \right)
	=
	&+ 2 \sum_{\vec{k'},s'} \text{Re} \left[ \g{k'}{s'}{\xi_{L}}
	\langle \bc{\xi_{L}} \vc{k'}{s'} \ca{k'}{s'} \rangle \right]
	- 2 \kappa_{\xi_{L}} \langle \bc{\xi_{L}} \ba{\xi_{L}} \rangle
\quad, \\
	\frac{\text{d}}{\text{d}t}
	\left( \langle \bc{\xi_{L}} \vc{k}{s} \ca{k}{s} \rangle \right)
	=
	&- \frac{i}{\hbar} \left( \eps{k}{s}{c} - \eps{k}{s}{v} - \hbar \omega_{\xi_{L}} \right)
	\langle \bc{\xi_{L}} \vc{k}{s} \ca{k}{s} \rangle
	\\
	&+ \gast{k}{s}{\xi_{L}}
	\left(
	\langle \cc{k}{s} \ca{k}{s} \rangle \left( 1 - \langle \vc{k}{s} \va{k}{s} \rangle \right)
	+ \langle \bc{\xi_{L}} \ba{\xi_{L}} \rangle
	\left( \langle \cc{k}{s} \ca{k}{s} \rangle - \langle \vc{k}{s} \va{k}{s} \rangle \right)
	\right)
	\\
	&+ \gast{k}{s}{\xi_{L}} \left( \delta \langle \bc{\xi_{L}} \ba{\xi_{L}} \cc{k}{s} \ca{k}{s} \rangle
	- \delta \langle \bc{\xi_{L}} \ba{\xi_{L}} \vc{k}{s} \va{k}{s} \rangle \right)
	\\
	&- \left(\kappa_{\xi_{L}} + \Gamma \right)
	\langle \bc{\xi_{L}} \vc{k}{s} \ca{k}{s} \rangle \quad.
\end{align*}

\section{$g^{(1)}(\tau)$ and coherence time $\tau_{c}$}

Access to the temporal evolution of the first-order correlation function $g^{(1)}(\tau)$ is required in order to calculate the
single-photon spectrum as well as the coherence time. \\
$g^{(1)}(\tau)$ describes the correlation of photons with themselves at different times $t_{s}$ and $t_{s} + \tau$, where $t_{s}$ is the time at which
the system has reached a stationary state; the function is normalised using the photon number for the stationary state and hence reads:
\cite[eq. (4.1)]{glauber1963quantum}

\begin{align*}
	g^{(1)}(\tau) = \frac{\langle \bc{\xi_{L}}(t_{s}) \ba{\xi_{L}}(t_{s} + \tau) \rangle}
	{\langle \bc{\xi_{L}}(t_{s}) \ba{\xi_{L}}(t_{s}) \rangle} \quad.
\end{align*}

The time over which the correlation of a single photon with its delayed self decays, defines the coherence time.
Essentially, it results as an integration of the one-photon correlation function over the delay time $\tau$:
\cite[eq. (3.4.13)]{loudon2000quantum}

\begin{align*}
	\tau_{c} = \int_{-\infty}^{+\infty} | g^{(1)}(\tau) |^2 \text{d}\tau \quad.
\end{align*}

In order to calculate the two-time expectation value $\langle \bc{\xi_{L}}(t_{s}) \ba{\xi_{L}}(t_{s} + \tau) \rangle$,
one first considers the dynamics of $\langle \ba{\xi_{L}} \rangle$ which can be calculated by means of the Heisenberg EoM including
Lindblad terms for the resonator and cavity decay losses:

\begin{align*}
	\frac{\text{d}}{\text{d} \tau'} \langle \ba{\xi_{L}} \rangle
	= - \left( \kappa_{\xi_{L}} + i \omega_{\xi_{L}} \right) \langle \ba{\xi_{L}} \rangle
	+ \sum_{\vec{k'},s'} \g{k'}{s'}{\xi_{L}} \langle \vc{k'}{s'} \ca{k'}{s'} \rangle \quad.
\end{align*}

\newpage

The quantum regression theorem (QRT) states that if a correlation function $\langle \hat{A}_{\mu}(t+\tau) \rangle$ follows certain dynamics as
a function of $\tau$, the same dynamics apply to those correlation functions defined via $\langle \hat{O}(t) \hat{A}_{\mu}(t+\tau) \rangle$
($\mu = 1,2,...$ and $\tau \geq 0$), where $\hat{O}$ is an arbitrary operator. \cite[p. 25/26]{carmichael1999statistical} \\
Introducing $\tau' \rightarrow t_{s} + \tau$ and a phase factor $e^{i \omega_{\xi_{L}} \tau}$,
followed by the application of the QRT, then leads to:

\begin{align*}
	\frac{\text{d}}{\text{d} \tau} \langle \bc{\xi_{L}}(t_{s}) \ba{\xi_{L}}(t_{s}+\tau) \rangle e^{i \omega_{\xi_{L}} \tau}
	= &\vphantom{+} \sum_{\vec{k'},s'} \g{k'}{s'}{\xi_{L}}
	\underbrace{\langle \bc{\xi_{L}}(t_{s}) \vc{k'}{s'}(t_{s}+\tau) \ca{k'}{s'}(t_{s}+\tau) \rangle e^{i \omega_{\xi_{L}} \tau}}_
	{\equiv P_{\vec{k'},s',\xi_{L}}(\tau)}
	\\
	&- \kappa_{\xi_{L}} \underbrace{\langle \bc{\xi_{L}}(t_{s}) \ba{\xi_{L}}(t_{s}+\tau) \rangle e^{i \omega_{\xi_{L}} \tau}}_
	{\equiv G_{\xi_{L}}(\tau)} \quad.
\end{align*}

Similarly, one can derive the Heisenberg EoM for $\langle \vc{k}{s} \ca{k}{s} \rangle$, introduce dephasing and the rotating frame
and apply the cluster expansion to the resulting $\langle \ba{\xi_{L}} \cc{k}{s} \ca{k}{s} \rangle$ and
$\langle \ba{\xi_{L}} \vc{k}{s} \va{k}{s} \rangle$ terms, neglecting higher order correlations.
Lastly, the QRT can be used to determine the dynamics of the polarisation defined at different times $t_{s}$ and $t_{s}+\tau$.

\begin{align*}
	\frac{\text{d}}{\text{d} \tau}
	\langle \bc{\xi_{L}}(t_{s}) \vc{k}{s}(t_{s}+\tau) \ca{k}{s}(t_{s}+\tau) \rangle e^{i \omega_{\xi_{L}} \tau}
	= &- \frac{i}{\hbar} \left( \eps{k}{s}{c} - \eps{k}{s}{v} - \hbar \omega_{\xi_{L}} + \Gamma \right)
	\underbrace{\langle \bc{\xi_{L}}(t_{s}) \vc{k'}{s'}(t_{s}+\tau) \ca{k'}{s'}(t_{s}+\tau) \rangle e^{i \omega_{\xi_{L}} \tau}}_
	{\equiv P_{\vec{k'},s',\xi_{L}}(\tau)}
	\\
	&+ \gast{k}{s}{\xi_{L}}
	\left( \langle \cc{k}{s}(t_{s}+\tau) \ca{k}{s}(t_{s}+\tau) \rangle
	- \langle \vc{k}{s}(t_{s}+\tau) \va{k}{s}(t_{s}+\tau) \rangle \right)
	\\
	&\times \underbrace{\langle \bc{\xi_{L}}(t_{s}) \ba{\xi_{L}}(t_{s}+\tau) \rangle e^{i \omega_{\xi_{L}} \tau}}_
	{\equiv G_{\xi_{L}}(\tau)} \quad.
\end{align*}

The derived quantity $G_{\xi_{L}}(\tau)$ can now be used to re-express $g^{(1)}(\tau)$ and thus $\tau_c$, respectively:

\begin{align*}
	\tau_{c} = \int_{-\infty}^{+\infty} | g^{(1)}(\tau) |^2 \text{d}\tau
	= 2 \int_{0}^{+\infty} \frac{|G_{\xi_{L}}(\tau)|^2}{|G_{\xi_{L}}(0)|^2} \text{d}\tau \quad.
\end{align*}

According to the Wiener-Khinchin theorem \cite[compare eq. (3.5.10)]{loudon2000quantum}, the Fourier transformation of
$g^{(1)}(\tau)$ then gives access to the spectrum, where the last identity follows per definition from $G(-\tau) = G(\tau)^{\ast}$
\cite[compare eq. (3.3.12)]{loudon2000quantum}:

\begin{align*}
	\mathcal{F} \left( g^{(1)}(\tau) \right)(\omega)
	= \int_{-\infty}^{+\infty} g^{(1)}(\tau) e^{i \omega \tau} \text{d}\tau
	= 2 \int_{0}^{+\infty} \text{Re}
	\left( \frac{G_{\xi_{L}}(\tau)}{G_{\xi_{L}}(0)} e^{i \left( \omega - \omega_{\xi_{L}} \right) \tau} \right) \text{d}\tau \quad.
\end{align*}

\newpage

\section{$g^{(2)}(\tau = 0)$ and quadruplet-level laser equations}

Even if the input-output characteristics as well as the coherence time coincide with the experimental data, one would still have to
investigate the zero-delay second-order correlation function since the photon statistics of coherent emission causes this quantity to
approach the Poisson limit with $g^{(2)}(\tau = 0) = 1$. \cite{jagsch2018quantum} \\
$g^{(2)}(\tau)$ describes the correlation of two photons with each other at different times $t_{s}$ and $t_{s} + \tau$, where $t_{s}$ is the time at
which the system has reached a stationary state; the function is normalised using the squared photon number for the stationary state and hence reads
for the zero-delay case: \cite[eq. (4.3)]{glauber1963quantum}

\begin{align*}
	g^{(2)}(\tau = 0)
	&= \frac{\langle \bc{\xi_{L}}(t_{s}) \bc{\xi_{L}}(t_{s}) \ba{\xi_{L}}(t_{s}) \ba{\xi_{L}}(t_{s}) \rangle}
	{\langle \bc{\xi_{L}}(t_{s}) \ba{\xi_{L}}(t_{s}) \rangle^2} \quad.
\end{align*}

The contributing higher-order correlations have to be calculated using the same approach of Heisenberg EoM
with Lindblad terms already applied when deriving the doublet-level equations. For this purpose, the following cluster expansions are used
which seize $\langle \bc{\xi} \bc{\xi} \rangle = \langle \ba{\xi} \ba{\xi} \rangle = 0$ (valid for quantities not coupling to the
populations driving the dynamics):

\begin{align*}
	\langle \bc{\xi_{L}} \bc{\xi_{L}} \ba{\xi_{L}} \ba{\xi_{L}} \rangle
	&= 2 \langle \bc{\xi_{L}} \ba{\xi_{L}} \rangle^2 + \delta \langle \bc{\xi_{L}} \bc{\xi_{L}} \ba{\xi_{L}} \ba{\xi_{L}} \rangle
\quad, \\
	\langle \bc{\xi_{L}} \bc{\xi_{L}} \ba{\xi_{L}} \vc{k}{s} \ca{k}{s} \rangle
	&= 2 \langle \bc{\xi_{L}} \ba{\xi_{L}} \rangle \langle \bc{\xi_{L}} \vc{k}{s} \ca{k}{s} \rangle
	+ \delta \langle \bc{\xi_{L}} \bc{\xi_{L}} \ba{\xi_{L}} \vc{k}{s} \ca{k}{s} \rangle \quad.
\end{align*}

With this, the Quantum Laser Equations on the quadruplet-level read:

\begin{align*}
	\frac{\text{d}}{\text{d}t} \left( \delta \langle \bc{\xi_{L}} \bc{\xi_{L}} \ba{\xi_{L}} \ba{\xi_{L}} \rangle \right)
	=
	&+ 4 \sum_{\vec{k'},s'} \text{Re} \left[ \g{k'}{s}{\xi_{L}}
	\delta \langle \bc{\xi_{L}} \bc{\xi_{L}} \ba{\xi_{L}} \ba{\xi_{L}} \rangle \right]
	- 4 \kappa_{\xi_{L}} \delta \langle \bc{\xi_{L}} \bc{\xi_{L}} \ba{\xi_{L}} \ba{\xi_{L}} \rangle
\quad, \\
	\frac{\text{d}}{\text{d}t} \left( \delta \langle \bc{\xi_{L}} \ba{\xi_{L}} \cc{k}{s} \ca{k}{s} \rangle \right)
	=
	&- 2 \text{Re} \left[ \g{k}{s}{\xi_{L}} \delta \langle \bc{\xi_{L}} \bc{\xi_{L}} \ba{\xi_{L}} \vc{k}{s} \ca{k}{s} \rangle
	+ \gast{k}{s}{\xi_{L}} \left( \langle \cc{k}{s} \ca{k}{s} \rangle + \langle \bc{\xi_{L}} \ba{\xi_{L}} \rangle \right)
	\langle \bc{\xi_{L}} \vc{k}{s} \ca{k}{s} \rangle \right]
	\\
	&- 2 \kappa_{\xi_{L}} \delta \langle \bc{\xi_{L}} \ba{\xi_{L}} \cc{k}{s} \ca{k}{s} \rangle
\quad, \\
	\frac{\text{d}}{\text{d}t} \left( \delta \langle \bc{\xi_{L}} \ba{\xi_{L}} \vc{k}{s} \va{k}{s}\rangle \right)
	=
	&+ 2 \text{Re} \left[ \g{k}{s}{\xi_{L}} \delta \langle \bc{\xi_{L}} \bc{\xi_{L}} \ba{\xi_{L}} \vc{k}{s} \ca{k}{s} \rangle
	+ \gast{k}{s}{\xi_{L}} \left( 1 - \langle \vc{k}{s} \va{k}{s} \rangle + \langle \bc{\xi_{L}} \ba{\xi_{L}} \rangle \right)
	\langle \bc{\xi_{L}} \vc{k}{s} \ca{k}{s} \rangle \right]
	\\
	&- 2 \kappa_{\xi_{L}} \delta \langle \bc{\xi_{L}} \ba{\xi_{L}} \cc{k}{s} \ca{k}{s} \rangle
\quad, \\
	\frac{\text{d}}{\text{d}t} \left( \delta \langle \bc{\xi_{L}} \bc{\xi_{L}} \ba{\xi_{L}} \vc{k}{s} \ca{k}{s} \rangle \right)
	=
	&- \frac{i}{\hbar} \left( \eps{k}{s}{c} - \eps{k}{s}{v} - \hbar \omega_{\xi_{L}} \right)
	\delta \langle \bc{\xi_{L}} \bc{\xi_{L}} \ba{\xi_{L}} \vc{k}{s} \ca{k}{s} \rangle
	\\
	&- 2 \g{k}{s}{\xi_{L}} \langle \bc{\xi_{L}} \vc{k}{s} \ca{k}{s} \rangle^2
	+ \left( \langle \vc{k}{s} \va{k}{s} \rangle - \langle \cc{k}{s} \ca{k}{s} \rangle \right)
	\delta \langle \bc{\xi_{L}} \bc{\xi_{L}} \ba{\xi_{L}} \ba{\xi_{L}} \rangle
	\\
	&+ 2 \g{k}{s}{\xi_{L}} \left(
	\left( 1 - \langle \vc{k}{s} \va{k}{s} \rangle + \langle \bc{\xi_{L}} \ba{\xi_{L}} \rangle \right)
	\delta \langle \bc{\xi_{L}} \ba{\xi_{L}} \cc{k}{s} \ca{k}{s} \rangle \right.
	\\
	&- \left. \left( \langle \cc{k}{s} \ca{k}{s} \rangle + \langle \bc{\xi_{L}} \ba{\xi_{L}} \rangle \right)
	\delta \langle \bc{\xi_{L}} \ba{\xi_{L}} \vc{k}{s} \va{k}{s} \rangle \right)
	\\
	&- \left(3 \kappa_{\xi_{L}} + \Gamma \right) \delta \langle \bc{\xi_{L}} \bc{\xi_{L}} \ba{\xi_{L}} \vc{k}{s} \ca{k}{s} \rangle \quad.
\end{align*}

\newpage

Using the cluster expansion for $\langle \bc{\xi_{L}} \bc{\xi_{L}} \ba{\xi_{L}} \ba{\xi_{L}} \rangle$ allows a straight-forward
definition of the two-photon correlation function $g^{(2)}(\tau = 0)$:

\begin{align*}
	g^{(2)}(\tau = 0)
	=
	2 + \frac{\delta \langle \bc{\xi_{L}}(t_{s}) \bc{\xi_{L}}(t_{s}) \ba{\xi_{L}}(t_{s}) \ba{\xi_{L}}(t_{s}) \rangle}
	{\langle \bc{\xi_{L}}(t_{s}) \ba{\xi_{L}}(t_{s}) \rangle^2} \quad.
\end{align*}

\section{Adiabatic elimination procedure for performance improvement}

A direct solution of the Quantum Laser Equations on the quadruplet level leads to a rather significant computational effort.
Since stationary solutions are of primary interest here, the adiabatic elimination of the photon-assisted polarisation
$\langle \bc{\xi_{L}} \vc{k}{s} \ca{k}{s} \rangle$ as well as the higher order polarisation-like quantity
$\delta \langle \bc{\xi_{L}} \bc{\xi_{L}} \ba{\xi_{L}} \vc{k}{s} \ca{k}{s} \rangle$ is a useful step in order to decrease the numerical cost
of executing these calculations, in particular when addressing a high resolution in terms of $|\vec{k}|$- as well as pump rate values. \\
Seeking for adiabatic solutions of polarisation-like terms leads to:

\begin{align*}
	\langle \bc{\xi_{L}} \vc{k}{s} \ca{k}{s} \rangle_{0}
	&=
	\frac{1}{\frac{i}{\hbar} \left( \eps{k}{s}{c} - \eps{k}{s}{v} - \hbar \omega_{\xi_{L}} \right) + \left(\kappa_{\xi_{L}} + \Gamma \right)}
	\\
	&\times \left( \gast{k}{s}{\xi_{L}} \left(
	\langle \cc{k}{s} \ca{k}{s} \rangle \left( 1 - \langle \vc{k}{s} \va{k}{s} \rangle \right)
	+ \langle \bc{\xi_{L}} \ba{\xi_{L}} \rangle
	\left( \langle \cc{k}{s} \ca{k}{s} \rangle - \langle \vc{k}{s} \va{k}{s} \rangle \right)
	\right) \right.
	\\
	&+ \left. \gast{k}{s}{\xi_{L}} \left( \delta \langle \bc{\xi_{L}} \ba{\xi_{L}} \cc{k}{s} \ca{k}{s} \rangle
	- \delta \langle \bc{\xi_{L}} \ba{\xi_{L}} \vc{k}{s} \va{k}{s} \rangle \right) \right)
\quad, \\
	\delta \langle \bc{\xi_{L}} \bc{\xi_{L}} \ba{\xi_{L}} \vc{k}{s} \ca{k}{s} \rangle_{0}
	&=
	\frac{1}{\frac{i}{\hbar} \left( \eps{k}{s}{c} - \eps{k}{s}{v} - \hbar \omega_{\xi_{L}} \right) + \left(3 \kappa_{\xi_{L}} + \Gamma \right)}
	\\
	&\times \left(- 2 \g{k}{s}{\xi_{L}} \langle \bc{\xi_{L}} \vc{k}{s} \ca{k}{s} \rangle^2_{0}
	+ \left( \langle \vc{k}{s} \va{k}{s} \rangle - \langle \cc{k}{s} \ca{k}{s} \rangle \right)
	\delta \langle \bc{\xi_{L}} \bc{\xi_{L}} \ba{\xi_{L}} \ba{\xi_{L}} \rangle \right.
	\\
	&+ \left. 2 \g{k}{s}{\xi_{L}} \left(
	\left( 1 - \langle \vc{k}{s} \va{k}{s} \rangle + \langle \bc{\xi_{L}} \ba{\xi_{L}} \rangle \right)
	\delta \langle \bc{\xi_{L}} \ba{\xi_{L}} \cc{k}{s} \ca{k}{s} \rangle \right. \right.
	\\
	&- \left. \left. \left( \langle \cc{k}{s} \ca{k}{s} \rangle + \langle \bc{\xi_{L}} \ba{\xi_{L}} \rangle \right)
	\delta \langle \bc{\xi_{L}} \ba{\xi_{L}} \vc{k}{s} \va{k}{s} \rangle \right) \right) \quad.
\end{align*}

Incorporating these adiabatically eliminated quantities as substitutions into the original dynamics also requires the calculation of their
corresponding real parts; in particular for the pre-factors this gives a Lorentzian function:

\begin{align*}
\text{Re} \left[
\frac{1}{\frac{i}{\hbar} \left( \eps{k}{s}{c} - \eps{k}{s}{v} - \hbar \omega_{\xi_{L}} \right) + \left(\alpha \kappa_{\xi_{L}} + \Gamma \right)}
\right]
=
\frac{\alpha \kappa_{\xi_{L}} + \Gamma}{\frac{1}{\hbar^2} \left( \eps{k}{s}{c} - \eps{k}{s}{v} - \hbar \omega_{\xi_{L}} \right)^2
+ \left(\alpha \kappa_{\xi_{L}} + \Gamma \right)^2} \quad.
\end{align*}

\newpage

In order to avoid artefacts of a constant dephasing, originating from an over-estimation of those states at large $|\vec{k}|$-values, we replace the
Lorentzian lineshape function with a hyperbolic secant function with more rapidly decreasing tails
\cite[compare eq. (2.68)]{chow1999semiconductor}, leading from

\begin{align*}
\frac{\alpha \kappa_{\xi_{L}} + \Gamma}{\frac{1}{\hbar^2} \left( \eps{k}{s}{c} - \eps{k}{s}{v} - \hbar \omega_{\xi_{L}} \right)^2
+ \left(\alpha \kappa_{\xi_{L}} + \Gamma \right)^2}
\end{align*}

to

\begin{align*}
\frac{1}{\alpha \kappa_{\xi_{L}} + \Gamma} \text{sech}
\left(
\frac{\frac{1}{\hbar} \left( \eps{k}{s}{c} - \eps{k}{s}{v} - \hbar \omega_{\xi_{L}} \right)}{\alpha \kappa_{\xi_{L}} + \Gamma}
\right) \quad.
\end{align*}

\section{Theoretical and experimental data in comparison}

For the comparison with experimental data, we use a single set of input-parameters given in Tab. S2.
Note that the results displayed below show three cases of different light-matter-coupling strengths, here referred to as "identical QWs" (red),
"inhom. case A" (blue) and "inhom. case B" (orange), the necessary details are also provided in the caption to Tab. S2.

\begin{table}[ht!]
\begin{center}
\begin{tabular}{ccc} \toprule
    {parameter} 		& {description} 						& {assigned value} 					\\	\toprule
    {$A$}				& {effective area}						& {$52500\,$nm$^2$} 				\\
   	{$g_{0}$}  			& {light-matter-coupling}  				& {$0.325\,$ps$^{-1}$}  					\\
    {$\gamma_{nl}$}  	& {spontaneous emission loss rate} 		& {$0.180\,$ps$^{-1}$} 					\\
    {$\kappa_{\xi_{L}}$}& {resonator loss / cavity decay rate}  & {$0.725\,$ps$^{-1}$} 					\\
  	{$\gamma_{rel}$}  	& {relaxation rate}  					& {$10.0\,$ps$^{-1}$} 					\\
    {$\Gamma$}  		& {dephasing}  							& {$5.0\,$ps$^{-1}$} 						\\ \bottomrule
\end{tabular}
\end{center}
\caption*{Tab. S2: Relevant simulation parameters employed for the comparison of experimental and theoretical data.
Note that in order to account for different light-matter-coupling strengths of individual quantum wells, $g_{0}$ has been varied
as follows: "ident." (6 x 100$\,\%$ of $g_{0}$), "inhom. case A" (1 x 25$\,\%$ of $g_{0}$, 1 x 50$\,\%$ of $g_{0}$, 1 x 75$\,\%$ of $g_{0}$,
1 x 125$\,\%$ of $g_{0}$, 1 x 150$\,\%$ of $g_{0}$, 1 x 175$\,\%$ of $g_{0}$), "inhom. case B" (1 x 80$\,\%$ of $g_{0}$,
1 x 90$\,\%$ of $g_{0}$, 1 x 100$\,\%$ of $g_{0}$, 1 x 120$\,\%$ of $g_{0}$, 1 x 140$\,\%$ of $g_{0}$, 1 x 160$\,\%$ of $g_{0}$).}
\end{table}

\newpage
\begin{figure}[ht!]
\begin{center}
{\makebox[\textwidth][c]{\fbox{\includegraphics[width=1.10\textwidth]{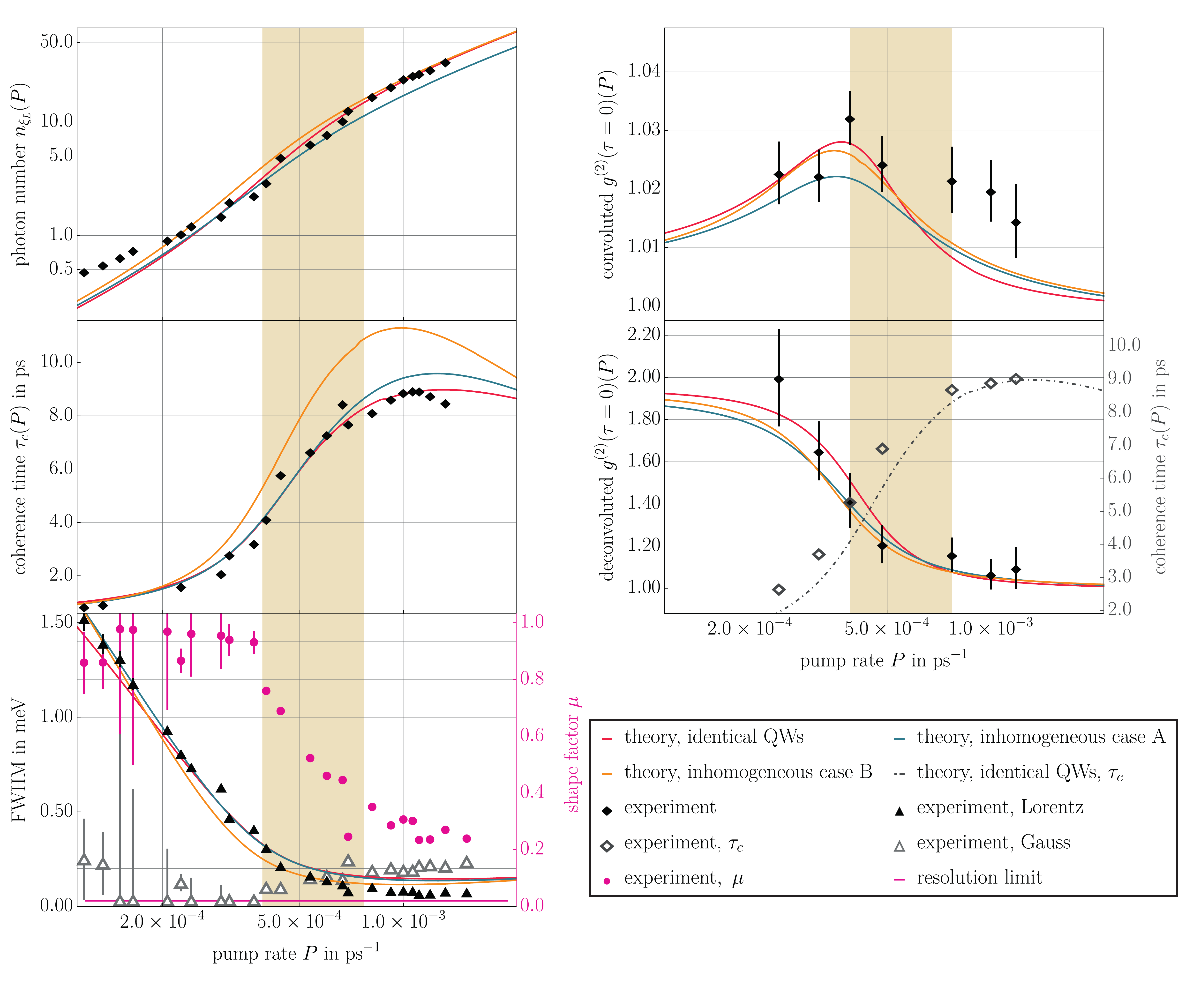}}}}
\caption*{Fig. S2: Excitation-power dependant characterisation of the MNL; the panels show (from top to bottom, left to right) a comparison of
experimental and theoretical data obtained from quantum-optical measurements and modelling, respectively. Namely: input-output characteristics,
coherence time, spectral Voigt analysis as well as the second-order photon-autocorrelation function $g^{(2)}(0)$. The theory curves show one scenario
for six identical (also presented in the main text) and two scenarios for six different light-matter interaction-strengths (governed by $g_{0,j}$ as
introduced in Sect. I) in order to investigate whether the Gaussian lineshape anomaly observed in the experiment may be due to the position-dependant
slightly varying overlap of the individual quantum wells' electronic wave functions and the lasing mode. However, the calculations presented here
indicate that this is not the intrinsic effect responsible for the observed anomaly.}
\end{center}
\end{figure}

\newpage

\section*{Semi-classical multimode laser model}
The semi-classical theoretical approach to investigate the emission spectrum of a semiconductor laser from below to above lasing threshold
uses composite/free-space eigenmodes given by:

\begin{align}\tag{1}
	\frac{\text{d}^2}{\text{d}z^2} u_{m}(z) = - \frac{n^2(z)}{c^2} \Omega^2_{m} u_{m}(z)
\end{align}

to account for outcoupling \cite{Chow2021, lang1973laser}. In Eq. (1), $c$ is the speed of light in vacuum, $\Omega_{m}$ is the passive cavity frequency and $z$
extends over the laser cavity and free space. The laser cavity and its outcoupling mirrors are described by the spatial dependence of the effective
refractive index $n(z)$ (as visualised in Fig. 5 of the paper). For each mode $u_{m}(z)$, the boundary conditions are the continuity of each
eigenfunction and first derivative at each refractive index interface. In this approach, every Fox-Li mode (customary used in the derivation of the
laser equations) comprises a large number of composite laser/free-space modes. Fox-Li modes decay because of outcoupling and thus do not form a strictly
orthonormal basis. In contrast, each composite laser/free-space mode rigorously satisfies orthogonality:

\begin{align}\tag{2}
	\int^L_{0} n^2(z) u_{n}(z) u_{m}(z) \text{d}z = \mathcal{N} \delta_{n,m}
\end{align}

with normalization $\mathcal{N} = \nicefrac{1}{2} \sum^{n_{\text{cav}}}_{j=1} n^2_{j} L_{j}$, where $L$ is the total cavity length, $L_j$ is the length of each section (laser cavity or free-space) and $n_j$ denotes the corresponding refractive index of the section.

~ \\

With the orthogonal composite laser/free-space basis, we write the laser field:

\begin{align}\tag{3}
	E(z,t) = \frac{1}{2} \sum_{n} E_{n}(t) e^{-i(\nu_{n}t + \phi_{n}(t))} u_{n}(z) + c.c.
\end{align}

Performing the usual derivation in a multimode laser theory \cite{sargent1974} gives the intensity- and frequency-determining equations:

\begin{align}\tag{4}
	\frac{\text{d} \mathcal{I}_{n}}{\text{d}t} =
	\left[ g^{\text{sat}}_{n}(N_{2d}) - \gamma^{\text{cav}}_{n} \right] \mathcal{I}_{n} + S_{n}(N_{2d})
	+ \sum_{m \neq n} 2 \sqrt{\mathcal{I}_{n} \mathcal{I}_{m}} \text{Re} \left[ B_{nm}(N_{2d}) e^{-i \psi_{nm}} \right]
\end{align}

\begin{align}\tag{5}
	\frac{\text{d} \psi_{n}}{\text{d}t} =
	\Omega_{n} + \left[ \sigma_{n}(N_{2d}) - \sum_{m} \tau_{nm}(N_{2d}) \mathcal{I}_{m} \right]
	- S_{\phi}(N_{2d}) + \sum_{m \neq n} \sqrt{\frac{\mathcal{I}_{m}}{\mathcal{I}_{n}}} \text{Im} \left[ B_{nm}(N_{2d}) e^{-i \psi_{nm}} \right]
\end{align}

where $\text{d}\psi_{n}/\text{d}t = \nu_{n} + \text{d}\phi_{n}/\text{d}t$ is the n$^{\text{th}}$ composite cavity mode lasing frequency , $\psi_{nm} = \psi_{n} - \psi_{m}$ and $\mathcal{I}_{n} = (\wp E_{n} / (2 \hbar \gamma))^2$.
$\wp$ and $\gamma$ are the dipole matrix element and dephasing rate, respectively. In the RHS of Eq. (4), the first two terms are the modal saturated
gain and cavity loss, the third term accounts for the spontaneous emission and the last term is from the first order polarization, arising because
the composite modes are not orthogonal when integrated over only the laser cavity. The square brackets in Eq. (5) contain the modification to the
passive composite mode frequency $\Omega_{n}$ by the active medium. They are the frequency pulling and pushing contributions $\sigma_{n}$ and
$\tau_{nm}$, respectively. In semiconductor laser models they arise from the carrier-induced refractive index change. The contribution $S^{\phi}_n$
accounts for phase diffusion from spontaneous emission. Coefficients associated with Eq.s (4) and (5) are derived from the electron-hole polarization
equation of motion \cite{chow2020multimode}. From a cavity-QED derivation, we obtain the equation of motion for a single-mode laser field. We extract from it
\cite{chow2018quantum,chow1975fluctuation}:

\begin{align}\tag{6}
	S_{n} = \frac{\varepsilon_{g_{0}} N_{\text{qw}} w L_{g}}{\epsilon_{B} V_{\text{mode}}}
	\left( \frac{\wp}{2 \hbar \gamma} \right)^2 \Gamma_{nn}^{(1)} \beta_{\text{spont}} B_{\text{spont}}^{(2d)} N_{2d}^2
	f(\varepsilon_{n}^{e}, \mu_{e}, T) f(\varepsilon_{n}^{h}, \mu_{h}, T) 
\end{align}

\begin{align}\tag{7}
	S_{\phi} = i \gamma_{n}^{\text{cav}}\frac{\hbar \nu}{2\epsilon_{B} V_{\text{mode}}} \left( \frac{\wp}{2 \hbar \gamma} \right)^{2} \frac{1}{\mathcal{I}_{n}} 
\end{align}

where $w$ and $L_{g}$ are the stripe width and length of the active region, $\Gamma_{nn}^{(1)}$ is the linear mode confinement factor in the active region, $\beta_{\text{spont}}$ is the spontaneous emission factor,
$f(\varepsilon_{n}^{e}, \mu_{e}, T)$ and $f(\varepsilon_{n}^{h}, \mu_{h}, T)$ are the electron and hole populations (assuming Fermi functions)
contributing to the emission into the $n$th composite cavity mode. \\

Owing to rapid carrier-carrier scattering, the intensity- and frequency-determining equations alone do not determine laser behavior. An expedient
approach to account for the scattering is to evaluate all active medium coefficients at the saturated carrier density $N_{2d}$, obtained by
simultaneously solving Eq.s (4) and (5) with the total carrier density equation of motion:

\begin{align}\tag{8}
	\frac{\text{d} N_{2d}}{\text{d}t} = - \frac{\epsilon_{B} h_{\text{qw}}}{8 \hbar \nu_{0}} \left( \frac{\wp}{2 \hbar \gamma} \right)^{-2}
	\frac{1}{\Gamma_{xy}} \sum_{n} g_{n}^{\text{sat}} \mathcal{I}_{n} + \frac{\eta_{p} J}{e N_{\text{qw}}} - \gamma_{\text{nr}} N_{2d}
	- B_{\text{spont}}^{(2d)} N_{2d}^2
\end{align}

where $\epsilon_{B}$ and $\nu_{0}$ are the averaged permittivity and frequency, $h_{\text{qw}}$ is the thickness of one of the quantum wells,
$\Gamma_{xy}$ is the transverse confinement factor, $\eta_{p}$ is the pump efficiency due to Pauli blocking, $J$ is the injection current density,
$N_{\text{qw}}$ is the number of quantum wells in the active medium, $\gamma_{\text{nr}}$ is the non-radiative (Shockley-Read-Hall) carrier loss and
$B_{\text{spont}}^{(2d)}$ is the bimolecular carrier recombination due to spontaneous emission. Finally, the lasing spectra (shown in Fig. S3) can be calculated according to the Wiener-Khinchin theorem \cite{loudon2000quantum}. The set of the employed input parameters is listed in Tab. S3. 

\begin{table}[ht!]
\begin{center}
\begin{tabular}{ccc}\toprule
    {parameter} 				& {description} 						& {assigned value} 					\\	\toprule
    {$m_e$}					& {electron effective mass}				& {$0.057\,m_{0}$} 				\\
    {$m_{h}$}	  			& {hole effective mass}  					& {$0.135\,m_{0}$ }					\\
    {$\varepsilon_{g}$}  			& {QW bandgap energy} 					& {$883\,$meV} 					\\
    {$\wp$}					& {dipole matrix element}  				& {$0.340\,$nm} 					\\
    {$n_{b}$}  				& {refractive index (high frequency)}  			& {$3.334$} 					\\
    {$n_{0}$}  				& {refractive index (low frequency)}  			& {$3.734$} 					\\
    {$N_{\text{qw}}$}  			& {QW layers}  						& {$6$} 					\\
    {$h_{\text{qw}}$}  			& {QW thickness}  						& {$6.0\,$nm} 					\\
    {$V_{\text{mod}}$}  			& {mode volume}  						& {$390\cdot 10^{6}\,$nm$^3$} 					\\
    {$\Gamma_{xy}$}  			& {transverse confinement factor}  			& {$0.055$} 							\\
    {$\beta$}  				& {spontaneous emission factor}  			& {$5.0\cdot 10^{-4}$} 							\\
    {$\gamma$}  				& {dephasing rate}  					& {$10.0\,$ps$^{-1}$} 						\\ 
    {$\gamma_{ab}$}  			& {population relaxation rate}  				& {$10.0\,$ps$^{-1}$} 						\\ 
    {$\gamma_{\text{nr}}$}  		& {nonradiative carrier loss rate}  			& {$5.0\cdot 10^{-4}\,$ps$^{-1}$} 						\\
    {$\alpha_{\text{abs}}$}  		& {absorption}  						& {$5.0\cdot10^{-7}\,$nm$^{-1}$} 						\\   
    {$B_{\text{spont}}^{(2d)}$}  	& {bimolecular carrier recombination rate}  		& {$0.004\,$nm$^{2}$ ps$^{-1}$} 				\\
    {$\gamma_{n}^{\text{cav}}$}  	& {cavity loss rate}  		          			& {$5.13\cdot10^{-4}\,$ps$^{-1}$} 		\\  
    {$\eta_{p}$}  			  	& {pump efficiency}  					& {$0.55\,$} \\ \bottomrule
\end{tabular}
\end{center}
\caption*{Tab. S3: Simulation parameters used in the composite laser/free-space semiclassical model.}
\end{table}

\newpage
\begin{figure}[ht!]
\begin{center}
{\makebox[\textwidth][c]{\fbox{\includegraphics[width=1.10\textwidth]{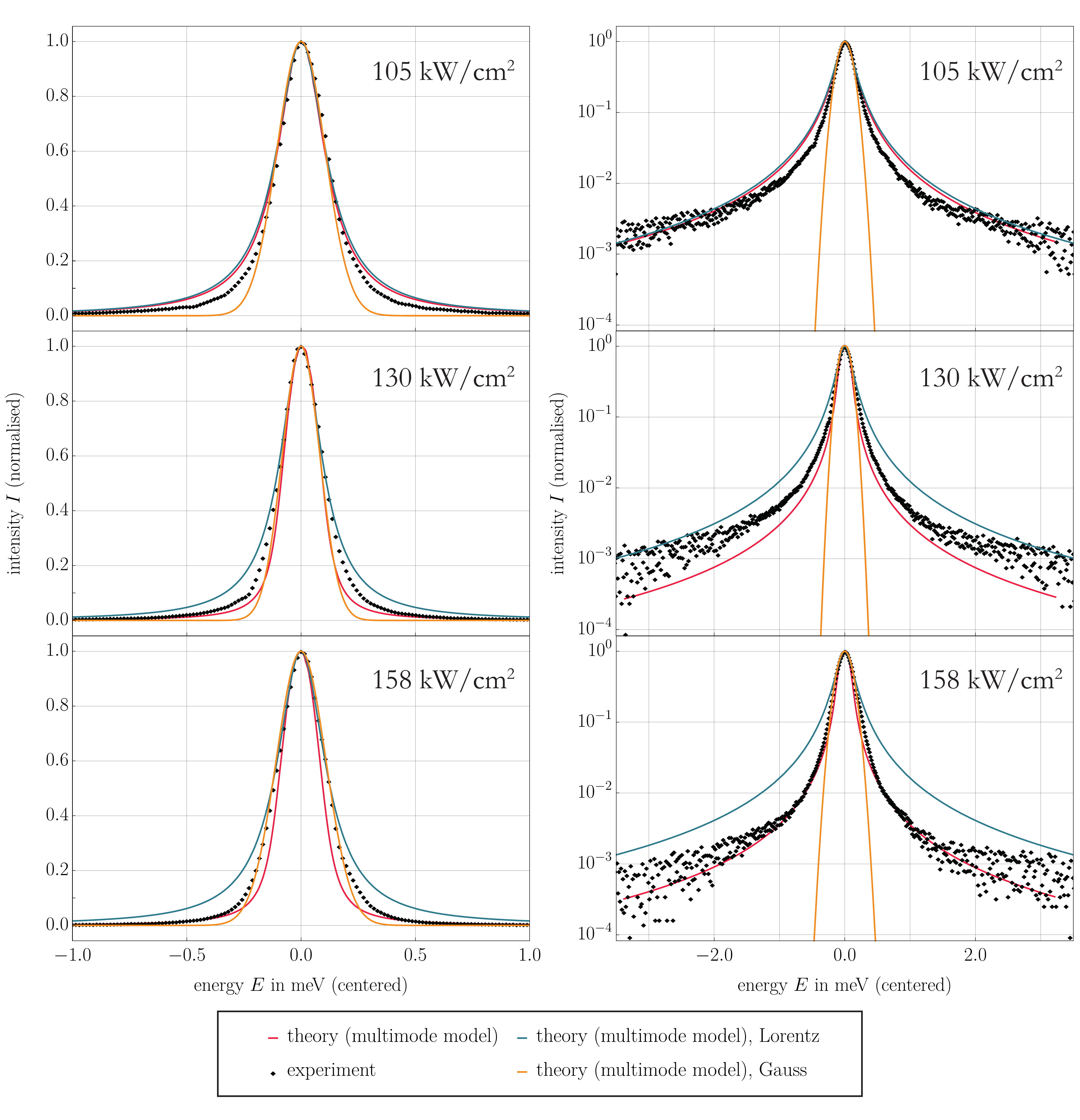}}}}
\caption*{Fig. S3: Emission spectra, obtained from the semiclassical model described here, at different excitation powers in comparison with the
experimental data (normal scale on the left side and logarithmic scale on the right side).
Shown are the lineshapes obtained from the theory (red) as well as the corresponding Lorentzian (blue) and Gaussian (orange) curves, experimental data
are displayed in black. It is evident that above the lasing threshold (around $65\,$kW/cm$^2$ or $3 \cdot 10^{-4}\,$ps$^{-1}$), the lineshape is no
longer described by a Lorentzian curve alone, but clearly has a Gaussian character as well. The linewidth contributions have been analysed using a full
Voigt fit, the results of which are shown in Figure 6 of the main text.}
\end{center}
\end{figure}
\clearpage

\bibliographystyle{hplain}
